\documentclass[conference]{IEEEtran}
\IEEEoverridecommandlockouts
\usepackage{cite}
\usepackage{amsmath,amssymb,amsfonts}
\usepackage{graphicx}
\usepackage{textcomp}
\usepackage{xcolor}
\usepackage{multirow}%
\usepackage{amsthm}%
\usepackage{mathrsfs}%
\usepackage{manyfoot}%
\usepackage{algorithm}%
\usepackage{algorithmicx}%
\usepackage[noend]{algpseudocode}%
\usepackage{listings}%
\usepackage{subcaption}
\usepackage[implicit=false,hidelinks,citecolor=.]{hyperref}
\usepackage{caption}
\makeatletter
\newlength{\continueindent}
\renewenvironment{algorithmic}[1][0]%
   {%
   \edef\ALG@numberfreq{#1}%
   \def\@currentlabel{\theALG@line}%
   \setcounter{ALG@line}{0}%
   \setcounter{ALG@rem}{0}%
   \let\\\algbreak%
   \expandafter\edef\csname ALG@currentblock@\theALG@nested\endcsname{0}%
   \expandafter\let\csname ALG@currentlifetime@\theALG@nested\endcsname\relax%
   \begin{list}%
      {\ALG@step}%
      {%
      \rightmargin\z@%
      \itemsep\z@ \itemindent\z@ \listparindent2em%
      \partopsep\z@ \parskip\z@ \parsep\z@%
      \labelsep 0.5em \topsep 0.2em%\skip 1.2em 
      \ifthenelse{\equal{#1}{0}}%
         {\labelwidth 0.5em}%
         {\labelwidth 1.2em}%
       \leftmargin\labelwidth \addtolength{\leftmargin}{\labelsep}
      \ALG@tlm\z@%
      }%
      \parshape 2 \leftmargin \linewidth \continueindent \dimexpr\linewidth-\continueindent\relax
   \setcounter{ALG@nested}{0}%
   \ALG@beginalgorithmic%
   }%
   {% end{algorithmic}
   \ALG@closeloops%
   \expandafter\ifnum\csname ALG@currentblock@\theALG@nested\endcsname=0\relax%
   \else%
      \PackageError{algorithmicx}{Some blocks are not closed!!!}{}%
   \fi%
   \ALG@endalgorithmic%
   \end{list}%
   }%
\makeatother
\algrenewcommand\algorithmicindent{0.5em}

\def\BibTeX{{\rm B\kern-.05em{\sc i\kern-.025em b}\kern-.08em
    T\kern-.1667em\lower.7ex\hbox{E}\kern-.125emX}}

\DeclareUnicodeCharacter{226A}{\ensuremath{\ll}}

\DeclareUnicodeCharacter{226B}{>>}
\usepackage{scalerel}

\begin{document}

\title{Edge-Mapping of Service Function Trees for Sensor Event Processing
\vspace{-0.3cm}%\\
}

% \IEEEauthorblockA{\textit{dept. name of organisation (of Aff.)} \\
% \textit{name of organisation (of Aff.)}\\
% \and
% \IEEEauthorblockA{\textit{dept. name of organisation (of Aff.)} \\
% \textit{name of organisation (of Aff.)}\\
% \and
% \IEEEauthorblockA{\textit{dept. name of organisation (of Aff.)} \\
% \textit{name of organisation (of Aff.)}\\
% \and
% \IEEEauthorblockA{\textit{dept. name of organisation (of Aff.)} \\
% \textit{name of organisation (of Aff.)}\\
% \and
% \IEEEauthorblockA{\textit{dept. name of organisation (of Aff.)} \\
% \textit{name of organisation (of Aff.)}\\
% \and
% \IEEEauthorblockA{\textit{dept. name of organisation (of Aff.)} \\
% \textit{name of organisation (of Aff.)}\\

\author{\IEEEauthorblockN{Babar Shahzaad,
Alistair Barros,
Colin Fidge
}

\IEEEauthorblockA{School of Information Systems,
Queensland University of Technology, Australia\\
\{babar.shahzaad, alistair.barros, c.fidge\}@qut.edu.au}
}

\maketitle

\begin{abstract}

Fog computing offers increased performance and efficiency for Industrial Internet of Things (IIoT) applications through distributed data processing in nearby proximity to sensors. Given resource constraints and their contentious use in IoT networks, current strategies strive to optimise which data processing tasks should be selected to run on fog devices. In this paper, we advance a more effective data processing architecture for optimisation purposes. Specifically, we consider the distinct functions of sensor data streaming, multi-stream data aggregation and event handling, required by IoT applications for identifying actionable events. We retrofit this event processing pipeline into a logical architecture, structured as a service function tree (SFT), comprising service function chains. We present a novel algorithm for mapping the SFT into a fog network topology in which nodes selected to process SFT functions (microservices) have the requisite resource capacity and network speed to meet their event processing deadlines. We used simulations to validate the algorithm's effectiveness in finding a successful SFT mapping to a physical network. Overall, our approach overcomes the bottlenecks of single service placement strategies for fog computing through composite service placements of SFTs.

%The Industrial Internet of Things (IIoT) necessitates solutions beyond cloud-based approaches due to real-time data processing demands. Fog computing with strategically placed fog nodes addresses this challenge. However, efficiently placing microservices on these fog nodes for processing sensor data streams remains challenging. We present a novel Service Function Tree (SFT) mapping algorithm for microservice placement in IIoT networks. Our new algorithm considers interdependencies of service function chains, service function order, fog device resource limitations, and sensor availability in regions of interest. Our new algorithm also employs backtracking and extended search strategies for mapping flexibility. We conduct simulations to validate the algorithm's effectiveness in finding successful SFT mapping to a physical network.

\end{abstract}


\section{Introduction}

%The Industrial Internet of Things (IIoT) envisions the seamless integration of various industrial devices equipped with sensing, processing, and actuation capabilities~\cite{xu2023survey}.
The Industrial Internet of Things (IIoT) is widely expected to automate cyber-physical domains, including manufacturing, transportation, healthcare, and public infrastructure through the integration of sensor devices and applications as well as enterprise systems and business processes~\cite{pivoto2021cyber}. Under a classical IoT model, data generated from sensors is streamed via gateways to central cloud systems for processing, event handling, and data analytics. The distance to the cloud and increasing data traffic in the core network can lead to significant transmission delays, thereby disrupting the strict  \textit{real-time data processing} requirement of many IoT applications~\cite{firouzi2022convergence}. Therefore, the conventional cloud is no longer suitable for such applications and not scalable for the \textit{business integration} requirements of the Industrial IoT~\cite{wu2020cloud,madakam2019industrial}.

With the increasing computational capacity of smart devices, \textit{fog computing}~\cite{chalapathi2021industrial} has emerged to support data processing tasks, typically in the form of microservices, on IIoT network nodes. This offloads some of the data processing to the cloud, reduces network transmission loads, and, crucially, detects and responds to events in close \textit{proximity to sensors}. Examples include \textit{predictive maintenance} in construction and manufacturing where equipment breakdowns and misplaced materials are detected and handled on-site, e.g., diagnosing faults and scheduling repairs or replacements~\cite{rao2022real}.

To support such IIoT scenarios, fog devices with suitable storage and computing capacity need to be strategically placed to cover physical work activities and sensing requirements. Moreover, \textit{limitations of network coverage} mean that not all devices are directly connected to each other and to sensors. Given concurrent sensing in dynamic IIoT settings, it is crucial to select which areas and sensors to target and which proximate resources to target for data processing. These selections should typically reflect areas of \textit{higher priority}, such as data bottlenecks or areas with increased occurrences of anomalies~\cite{qayyum2022mobility}. Data processing tasks are supported through microservices and strategies for addressing microservice-to-resource assignments include \textit{service placement optimisation}~\cite{pallewatta2022qos,barros2007multi}.

A major limitation of current service placement strategies is their focus on \textit{singleton tasks}. However, wider requirements for IIoT event processing include data streaming from sensors, filtering and aggregating data from different streams, and event handling through event-condition-action rule chains. With complex IIoT applications having thousands of event processing goals, a corresponding number of event processing pipelines need to be considered for service placements on fog devices. Specifically, for a given event processing goal, the microservices supporting different processing tasks~\cite{faticanti2020throughput,barros2000processes}, together with their dependencies, need to be mapped to proximate sensors, adequately resourced, and directly or indirectly connected to fog nodes. The fog nodes chosen must reflect a \textit{priority region of interest (ROI)} for event processing goals and the selected mappings must meet \textit{time-bounded} processing requirements (deadlines).

Rather than placing microservices individually, we reformulate the problem of microservice placement in the physical network as mapping an \textit{SFT} onto a substrate physical network.
%In this context, a physical network acts as a substrate network.
Existing studies typically use the term SFT to describe a multicast routing path for a group of sequential Service Function Chains (SFCs) originating from the same source node and belonging to the same kind of SFC~\cite{9967766}. An SFC is an ordered sequence of microservice functions for \textit{end-to-end service delivery} (e.g., event detection)~\cite{chen2022sfc}. SFCs are generally classified into three basic topologies~\cite{8847151}: linear, split-and-join, and open with branches, each originating from a single source and ending at single or multiple destinations. In the context of this paper, we use the term \textit{SFT to denote a hierarchical structure of interconnected SFCs configured as a tree}, where each SFC has its data source. The mapping of such an SFT onto a physical network of fog devices is constrained by \textit{interdependency relationships of SFCs}, \textit{the mapping of service functions in a specific order}, and \textit{the available resources of fog devices}. This mapping process is further complicated by the need to map microservices in the SFT onto only those devices that have the \textit{sensors required} by each microservice in its ROI.

Here we present a novel \textit{service function tree mapping algorithm} which finds a valid mapping of an SFT onto a physical network (if one exists). The algorithm considers the constraints above and recursively maps each microservice in the SFT to a suitable fog device in the physical network. In some cases, selecting a fog device to map a particular microservice in the SFT may lead to a dead-end for subsequent microservices, so our algorithm employs a backtracking strategy to explore alternative fog devices. Also, there may be situations where a fog device has the sensors required by a particular microservice in its communication range but cannot host the microservice due to resource, connectivity, or link speed constraints. In such situations, our algorithm performs an extended search to map the microservice to a suitable fog device while respecting all the constraints. To the best of our knowledge, no similar approach exists that takes into account the constraints mentioned above. Therefore, to validate our algorithm, we use simple to complex scenarios, demonstrating its effectiveness in finding valid mappings of SFTs to physical networks. The contributions of this paper are as follows:

\begin{itemize}
    \item  We reformulate the problem of microservice placement to a physical network as an SFT mapping problem.
    \item We present an SFT mapping algorithm that finds valid microservice placements onto a physical network.
\end{itemize}

\section{Concrete Pouring Scenario}

As motivation, we leverage real-time event processing in an \textit{Industry 4.0 scenario}~\cite{turner2020utilizing}, focusing on the concrete pouring process at a large IIoT site. The \textit{dynamic physical environment} of an IIoT site, with numerous data streams generated by the movement of machinery and workers, presents \textit{challenges} for data management and processing~\cite{raptis2019data}. These challenges become more complex during the concreting phase, which is \textit{sensitive} to external influences such as weather conditions~\cite{farzampour2019compressive}. The scenario involves a large concrete slab, forming an industrial building's floor, with a surface area of 400 square meters (sqm). The concrete pouring process is divided into \textit{sequential} pours of 10 sqm sections using mobile pouring troughs, covering the entire 400 sqm area. During the pouring process, workers constantly move pouring troughs and smooth the poured concrete to achieve a level surface. This back-and-forth movement, along with the curing process, is estimated to take approximately two hours to complete for the entire slab.

\begin{figure} [!htb]
    \centering
    \includegraphics[width=0.8\linewidth]{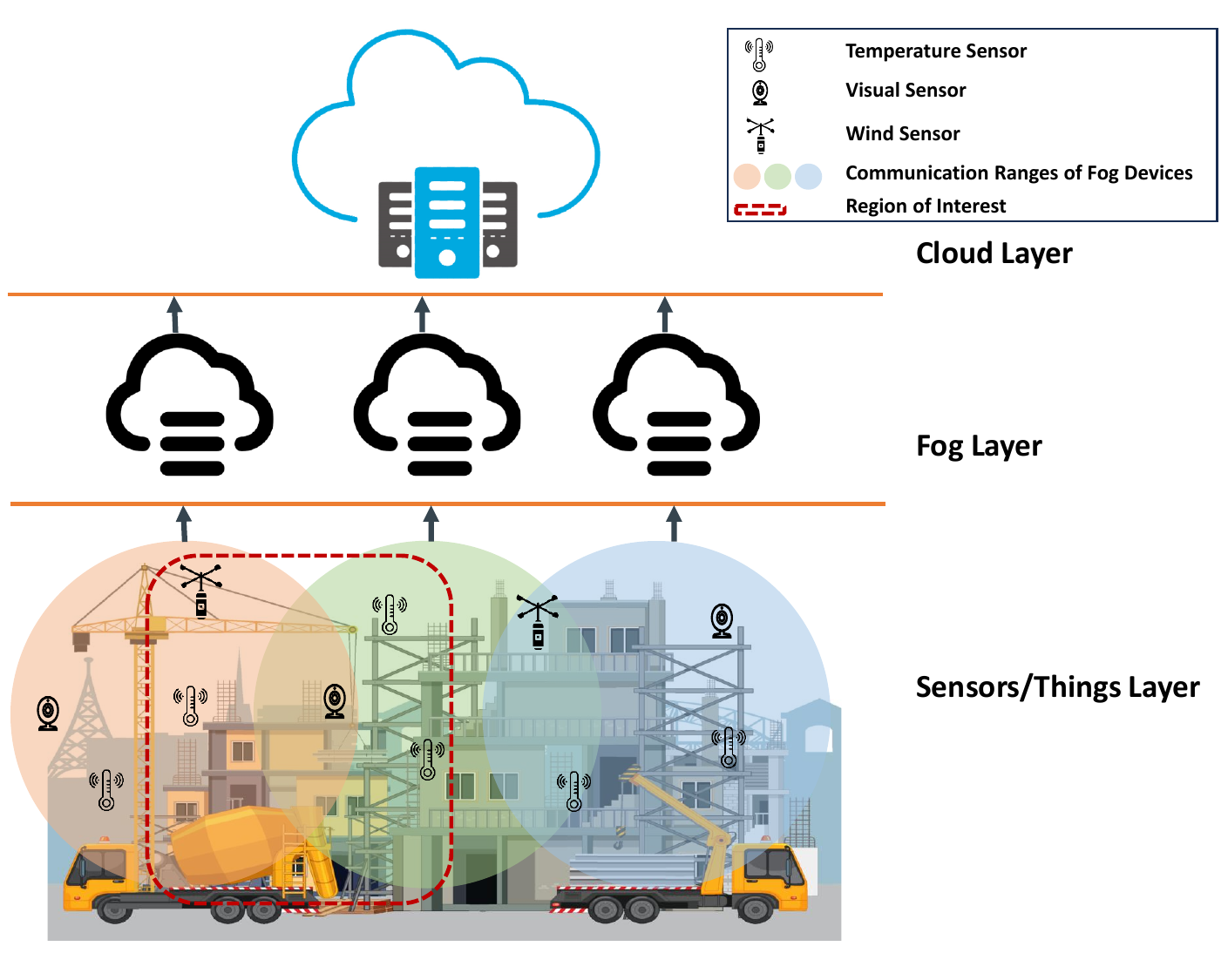}
    \caption{Fog-Cloud Architecture}
    \label{Sensor-Fog-Cloud}
\end{figure}

Various types of sensors are strategically deployed within the concrete pouring area to ensure concrete quality~\cite{arabshahi2021review}. The selection includes Geokon GK-8T temperature sensors for monitoring the temperature of the concrete mix, Honeywell HIH6130 humidity sensors for measuring water vapour, Sentek EC5 moisture sensors for tracking moisture content, Kyowa Electronic Instruments FLA-3-120-11 strain gauges for observing rebar stress, and Physical Acoustics PCI-II acoustic emission sensors for crack detection. These sensors are embedded within the concrete structure to provide real-time data critical for identifying events, such as rebar displacement, which can significantly impact the concrete's strength. Environmental factors and structural problems with rebars can affect the concrete~\cite{afzal2020reinforced}, leading to weaknesses that may take time to be visible. High temperatures or moisture levels during curing can lead to rapid setting or cracking. Wind can cause uneven drying, creating vulnerabilities. Immediate attention to these events, guided by sensor data, is crucial to mitigate structural vulnerabilities and ensure concrete quality.

The volume of data produced at IIoT sites poses a significant challenge for traditional data processing approaches, which may not be equipped to handle real-time data at a large scale. Processing all data from sensors in real time becomes resource-intensive and inefficient. To address this challenge, we introduce the concept of an ROI representing a specific area where focused attention is directed, especially in response to high-burst events (sudden temperature shifts) or anomalies (rebar displacement). The ROI's location and size vary depending on the number of data streams, e.g., an ROI covering a specific area of four pouring troughs covers 40 sqm. Identifying ROIs allows for strategic allocation of processing power and sensor data analysis. This ensures a targeted and efficient response to high-priority events and anomalies.

Data from ROIs is typically handled by microservices functioning as \emph{filters}, \emph{aggregators}, and \emph{event handlers}. They are organised in a hierarchical structure called a Service Function Tree (SFT), where nodes represent the microservices required for processing ROI sensor data allowing for an organised and efficient data processing workflow. For instance, temperature and humidity sensor data can be analysed to identify rapid drying conditions that could lead to cracks. Fog computing is crucial for ROI-based processing, as it facilitates near-real-time analysis by processing data closer to its source~\cite{pallewatta2023placement,decker2008non}. We employ Raspberry Pi~4 microcomputers as our fog computing devices.
% CF: I didn't see any value in this sentence.
%However, exploring fog computing devices with higher processing capabilities might be necessary for large-scale deployments with extensive sensor data.
This computing paradigm significantly reduces latency and the load on centralised cloud systems, ensuring timely and efficient data analysis. Fig.~\ref{Sensor-Fog-Cloud} illustrates the data flow from sensors in the ROI through the fog layer for initial processing and then to the cloud for further analysis.

\section{Related Work}

% To the best of our knowledge, no existing literature directly addresses SFT mapping for the sensor event processing within Fog-Cloud architectures. This study combines concepts from two separate areas: (1) Fog-Cloud Architecture and (2) Placement of SFTs in the Physical Network of Fog Devices. In this section, we overview related work in these two areas.

Previously, a Fog–Cloud Clustering (FCC) problem was presented as a multi-objective optimisation challenge aimed at enhancing resource management within Fog-to-Cloud (F2C) architectures~\cite{asensio2020designing}. It was modelled using a mixed-integer linear programming (MILP) formulation that considers various factors, such as minimising latency, reducing transmission power, and prioritising static nodes for leadership roles. The objective was to cluster fog devices in F2C scenarios using a machine learning-based heuristic to provide scalable near-optimal solutions. A significant assumption was the feasibility of the MILP model in large-scale IoT environments. This assumption was paramount because it addressed the computational constraints of MILP using a machine learning algorithm. Simulation experiments demonstrated the effectiveness of the proposed approach in providing solutions close to the MILP model with significantly reduced computational demands. However, this previously proposed approach did not consider how to focus on specific ROIs to access particular sensors' data and to optimise the data processing pipeline.
     
Elsewhere an IoT-driven service provisioning framework was proposed for efficient management of data streams from multiple IoT sources~\cite{li2021service}. It included mechanisms for uploading data streams, data aggregation and routing, and Virtual Network Functions (VNFs) placement. These data streams require a sequence of service functions to be executed in a specified order, forming an SFC. A heuristic algorithm was developed to minimise service response latency and operational costs of IoT-driven services. The algorithm leveraged the proposed service framework, focusing on optimising resource allocation and use within the mobile edge computing network to meet the SFC's requirements of IoT applications. However, this prior framework did not address SFC deployment based on sensor availability in the ROI.

Two service placement algorithms were also proposed to enhance the efficiency of network services in 5G networks through Network Function (NF) virtualisation~\cite{spinnewyn2018coordinated}. They addressed the challenge of optimally placing service requests in a physical infrastructure where NFs are subjected to location constraints due to various factors, such as latency, legislation, or hardware requirements. An Integer Linear Programming (ILP) formulation was used to find an optimal combination of service embedding and chain composition. Considering location constraints, this formulation provided an optimal placement of virtualised network functions and their interconnections. A greedy chain selection-based heuristic was proposed to solve the ILP for large-scale problems. It iteratively adds virtual NF chains to the service composition and embeds them simultaneously to find a near-optimal solution within a significantly reduced computation time. However, it was not considered that SFT's leaf nodes need to be mapped to physical nodes with the required sensors in range. Furthermore, the existence of direct or indirect paths and capacity requirements between mapped nodes were also not considered.

A two-stage algorithm was also proposed for embedding an SFT with minimum cost for Network Function Virtualisation (NFV)-enabled multicast~\cite{ren2019embedding}. It optimises the SFC embedding into a substrate network to minimise the overall setup cost of the SFC and the link connection costs towards clients. The first stage generates an initial feasible solution by embedding an SFC into the network and constructing a Steiner tree. The second stage optimises the initial solution by adding new instances of VNFs to transform the embedded SFC into an SFT, to further reduce delivery costs. The approach guarantees the best approximation ratio for solving the Steiner tree problem, indicating the efficiency and effectiveness of the approach. However, the algorithm does not consider the sensors' ranges, link speeds, and connectivity constraints for the SFT embedding to physical networks.

A Branching-Aware Service Function Graph (B\_SFG) problem was modelled to handle the complexities of branching traffic flows in NFV environments~\cite{jalalitabar2019branching}. It characterises the SFC as a more complex mesh-like SFG rather than a simple linear sequence of VNFs. A B\_SFG with a coordinated mapping algorithm was proposed to optimise the placement of SFCs and traffic routing in physical networks. It considers various constraints, including function dependencies and branching requirements, and uses dependency sorting and layering techniques to map service requests onto the physical network efficiently. However, it was assumed that nodes and links in the physical network have unlimited computing resources and bandwidth, which may not reflect real-world network constraints. Additionally, a framework of policies was defined to represent branching behaviour, execution order, and merging points for traffic flows within SFCs. Again, the approach did not address sensor range constraints, where an SFT node's placement requires proximity to certain sensors.

In other work, an ILP model and an approximation algorithm were proposed to embed aggregated service function trees (ASFT) in networks~\cite{guo2022optimal}. An ASFT merges multiple independent SFCs that cater to different flows into a single structure but require the same network security services. The ILP model focused on the optimal deployment of these ASFTs to minimise VNF setup and flow routing costs. A key challenge addressed was the efficient sharing of VNF instances among multiple SFCs to reduce redundancy and overall costs while ensuring that network security requirements are met. The approximation algorithm uses a maximisation submodular function and iteratively builds a solution by selecting the options that offer the greatest improvement to the objective function at each step. The approach reduces costs compared to existing methods in the context of NFV for enhancing network security. However, it mainly focuses on cost reduction through SFC consolidation instead of sensor data acquisition and capacity requirements between mapped nodes.

Most recently, a Service Chain Graph Design and Mapping (SCGDM) problem was presented in the context of NFV-enabled networks~\cite{he2022joint}. It was formulated as an ILP model to address interdependency constraints between various VNFs and dynamic changes in traffic volume. It aimed to minimise the maximum link load factor to enhance the overall network system performance. An approximation algorithm employing a randomised rounding method was proposed to provide a feasible solution within reasonable computational effort. The algorithm optimises both the design of service chain graphs (SCGs) and their mapping onto physical network resources.
% Extensive simulations are conducted to evaluate the proposed algorithm's performance, particularly in terms of resource utilisation and service latency.
However, the proposed SCGDM problem does not consider the ROI constraint for SCG's mapping in the physical network or the link requirements for hosting successive VNFs in an SCG.

Thus, to the best of our knowledge, existing research primarily focuses on deploying singular linear SFCs, with some studies exploring the deployment of composite SFTs. However, none of these prior studies incorporates sensor availability within ROIs into the service deployment process. Our research is the first to produce an SFT mapping algorithm that incorporates sensor selection, resource requirements, connectivity, and link speed constraints in the mapping process.

\section{Preliminaries}

This section describes our physical network and SFT models and then defines the constraint-based SFT mapping problem.

\subsection{Physical Network Model}

\begin{figure} [t]
    % trim=left bottom right top
    \centering
    \includegraphics[width=0.9\linewidth, trim = 5.2cm 2.6cm 7.3cm 6cm, clip]{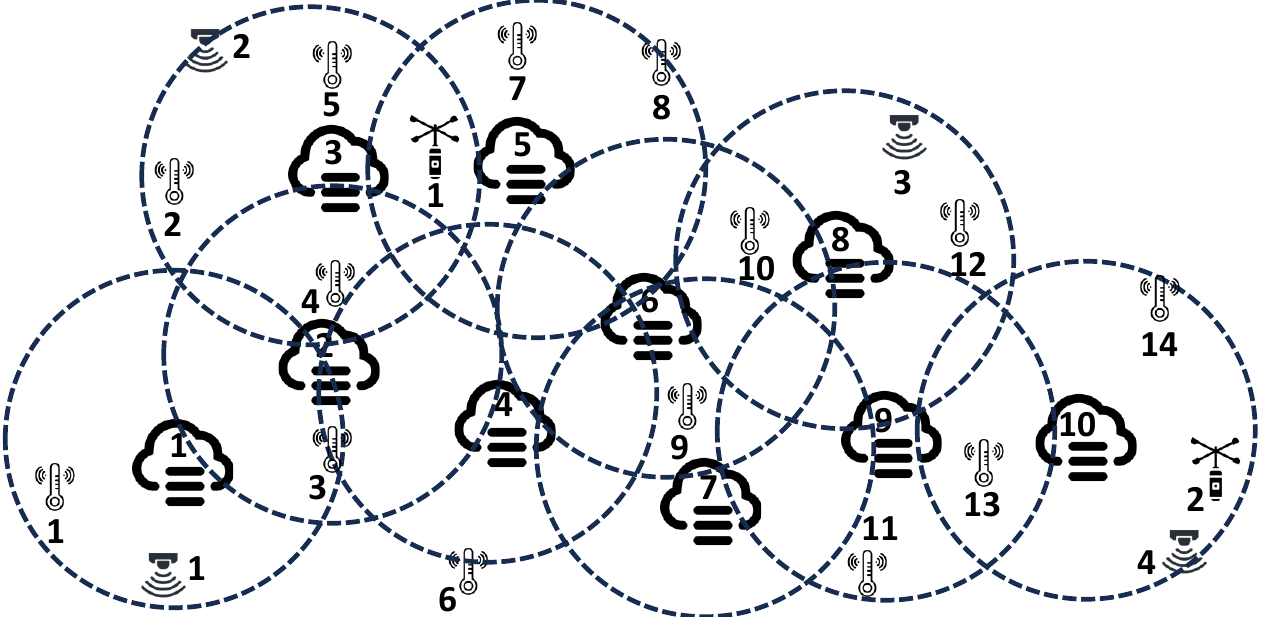}
    \caption{Physical Network in an IIoT Site}
    \label{PhysicalNetwork}
\end{figure}

Consider an IIoT environment that comprises physical sensors and fog devices. Let $S = \{s_{1}, s_{2}, \ldots, s_{n}\}$ be a set of $n$ sensors, where each sensor $s_{i}$ has an associated modality $\mu (s_i)$. Modalities represent the type of data that each sensor captures, such as temperature, vision, wind, structure, etc. Let \(D = \{d_1, d_2, \ldots, d_p\}\) be a set of $p$ fog devices. Each fog device has a subset of sensors in its communication range for data acquisition. The fog devices form a physical network to communicate with other fog devices in their range. We model this physical network (i.e., substrate network) as an undirected graph $PN = (D, L)$. Each node represents a fog device $d_i \in D$ and each edge represents a communication link between any pair of fog devices $(d_i, d_j) \in L$. Sensors located within the communication range of each fog device are represented as the attributes of that node (i.e., fog device) in the physical network. Each fog device $d_i \in D$ has a processing capacity $C(d_i)$ and each link $(d_i, d_j) \in L$ has a maximum transmission capacity of $C(d_i, d_j)$. In our physical network model, fog devices and sensors are statically placed within the IIoT site, and their locations are known a priori. These devices and sensors are strategically placed, leveraging wireless communication technologies like Zigbee and LoRa to observe complex operations, such as concrete pouring. Fig.~\ref{PhysicalNetwork} shows an example physical network in an IIoT site.
% It represents the physical network of sensors and fog devices essential for real-time event detection in an IIoT site environment.

\subsection{Service Function Tree Model}

We introduce an SFT model to represent a network of sensors and microservices that are required to process the sensors' data. As described earlier, an SFT is a composition of multiple interconnected Service Function Chains (SFCs). An SFC comprises a sequence of one or more microservice nodes that exchange data across connectivity links, where each service function represents the function of a specific microservice. We model the SFT as a directed acyclic graph: $SFT = (M, L')$, where $M = \{m_1, m_2, \ldots, m_p\}$ is a set of microservice nodes and $L'$ is the set of directed edges representing \textit{functional dependencies} between microservices.

\begin{figure}[t]
     \centering
     \begin{subfigure}[b]{0.1\textwidth}
         \centering
         \includegraphics[width=\linewidth, trim = 12.5cm 10.2cm 15.5cm 1cm, clip]{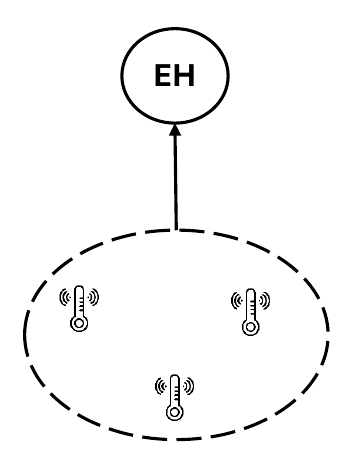}
         \caption{}
         \label{fig:et_5w}
     \end{subfigure}
     \hfill
      \begin{subfigure}[b]{0.1\textwidth}
         \centering
         \includegraphics[width=\linewidth, trim = 12.5cm 6.5cm 15.5cm 1cm, clip]{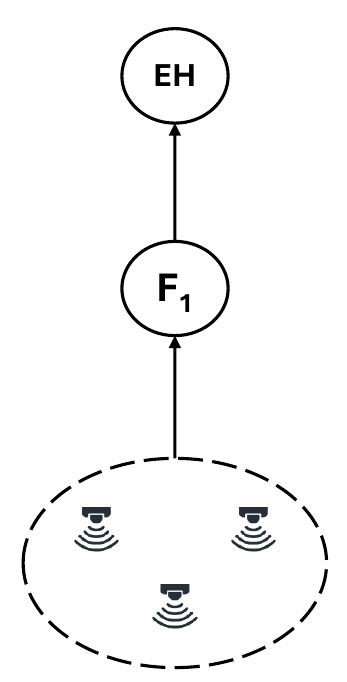}
         \caption{}
         \label{fig:et_4w}
     \end{subfigure}
     \hfill
      \begin{subfigure}[b]{0.25\textwidth}
         \centering
         \includegraphics[width=\linewidth, trim = 8.5cm 2cm 10.5cm 1cm, clip]{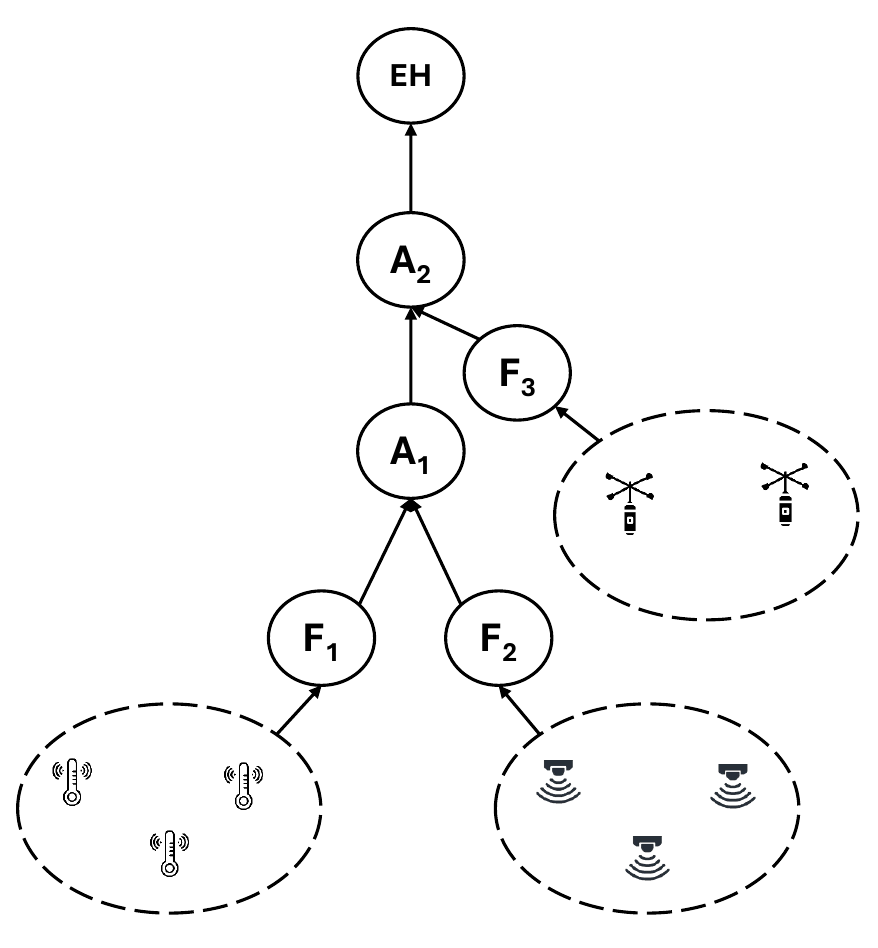}
          \caption{}
         \label{fig:et_3w}
     \end{subfigure}
     \hfill
     \caption{Service Function Tree Designs in an IIoT Site Scenario}
     \label{fig:sft}
\end{figure}

In an SFT, we identify three types of nodes: (1)~nodes with a single input and single output, (2)~nodes with multiple inputs and a single output, and (3)~nodes with single or multiple inputs but no output (apart from triggering actions not modelled here). Correspondingly, we categorise the microservices into three main types: Filters $F_i$, Aggregators $A_i$, and Event Handlers $EH$. Filters process and forward sensor data, aggregators combine data from multiple sources, and event handlers analyse incoming data to trigger actions and responses. Each microservice in the SFT has varying resource requirements \( R(m_i) \), which can be represented numerically as a specific value or a range, e.g., a memory requirement in GBs. Similarly, each connectivity link $(m_i, m_j) \in L'$ has communication requirements $R(m_i, m_j)$, which can represented numerically, e.g., a data transfer rate in Gbps. Leaf nodes in the SFT always have access to a predefined set of sensors within their ROI, from which they can elect to receive data. The SFT's root node is always an $EH$ microservice. We assume that the SFTs, mirroring a workflow process, are designed by an expert designer to reflect the necessary data processing strategy required in the IIoT site environment. Fig.~\ref{fig:sft} illustrates various example SFTs in the context of an IIoT site, e.g., Fig.~\ref{fig:et_3w} shows three SFCs forming an SFT as follows: $SFC_1 = \{F_1, A_1, A_2, EH\}$, $SFC_2 = \{F_2, A_1, A_2, EH\}$, and $SFC_3 = \{F_3, A_2, EH\}$.

\subsection{Assumptions}

Some underlying assumptions about the SFT and physical environment of the IIoT site are as follows:

\begin{itemize}
    \item The network's physical topology remains unchanged during the IIoT network's deployment.
    % \item Sensors and fog devices are strategically distributed across the IIoT site, and the sensors' location is known to fog devices.
    % \item Sensors are of various types, including temperature, visual, and wind sensors.
    \item Sensors and fog devices are assumed to have sufficient energy for the operational lifespan of their deployment.
    \item Radio interference and signal attenuation caused by the existence of physical objects are not taken into account.
    \item Each sensor collects data at a constant frequency and continuously transmits it to the fog devices in its range.
    \item The communication range of the fog devices is twice the sensing range of the sensors.
    % \item Each microservice in the SFT is designed to handle a singular data stream.
\end{itemize}

\subsection{Problem Formulation}

The problem of mapping an SFT onto a physical network is as follows: \textit{Given a physical network $PN$ and a service function tree $SFT$, the goal is to find a valid mapping of the SFT's nodes to physical network nodes while ensuring that the resource requirements are respected for each SFC in the SFT.}

We define a set of constraints for mapping an SFT of microservices onto a physical network of devices as follows:

\begin{itemize}
    \item \textbf{Sensor Selection Constraint.} We denote the sensors by unique identifiers within their respective regions of interest (ROIs) in an SFT. Sensors within each ROI share the same modality (i.e., sensor type). For a given microservice type \( m_i \in M \), only sensors of a specific modality from the designated ROI for $m_i$ can be chosen. We denote this constraint as follows:
    \begin{equation*}
    \scaleto{
        S(m_i) = \{s_i \in S \,|\, (\exists r \in ROI : (s_i, r) \in SR) \land \mu (s_i) = \mu (m_i)\}
    }{8.5pt}
    \end{equation*}
    % \normalsize
    where \(S\) is the set of all sensors with their associated modalities, \(ROI\) is the set of all regions of interest, \(SR\) is a set of pairs \((s_i, r)\) indicating that sensor \(s_i\) is located within region \(r\), $\mu (s_i)$ is the modality of sensor $s_i$, $\mu (m_i)$ is the required modality for microservice $m_i$, and \(S(m_i)\) is the function that returns the subset of sensors from which \(m_i\) can receive data.
    
    \item \textbf{Microservice Mapping Constraint.} Each microservice node $m_i \in M$ from the SFT (representing a node type \( F_i\), \(A_i, \) or \( EH \)) can be mapped to any fog device \( d_i \in D \) in the physical network \( PN \). We denote this as a mapping function \(map: SFT \rightarrow PN\).
    \item \textbf{Resource Allocation Constraint.} The sum of the resource requirements of the microservices allocated to any fog device \( d_i \in D \) must not exceed its capacity. We represent this constraint as follows:
   \[
   \forall d_i \in D, \sum_{m_i \in H_{d_i}} R(m_i) \leq C(d_i)
   \]
   where \(H_{d_i}\) is the set of microservices hosted in a fog device \( d_i \), \(R(m_i)\) represents the resource requirement of a microservice \(m_i\), and $C(d_i)$ denotes the resource capacity of the fog device $d_i$. In these constraints, we treat Boolean-valued expressions as returning 1 for True and 0 for False.
   \item \textbf{Path Connectivity Constraint.} For any two microservices \( m_i \) and \( m_j \) that are directly connected in the SFT, there must be a corresponding direct or indirect path between the fog devices $d_i$ and $d_j$ to which \( m_i \) and \( m_j \) are respectively mapped in the physical network $PN$. This is represented as:
   \[
   \forall (m_i, m_j) \in L', \exists \text{ a path between } d_i \text{ and } d_j \text{ in } PN
   \]
   \item \textbf{Link Capacity Constraint.} The total link speed requirements of all microservices mapped to a specific link in the physical network must not exceed the physical link's maximum transmission capacity. This constraint ensures efficient network usage and prevents network overloading. We represent this constraint as:
   \begin{equation*}
    \forall (d_i, d_j) \in L, \sum_{(m_i, m_j) \in L'_{(d_i, d_j)}} R(m_i, m_j) \leq C(d_i, d_j)
    \end{equation*}
   % \begin{equation*}
   % % \small
   %  \begin{split}
   %  \forall (d_i, d_j) \in L,
   %  \begin{cases}
   %  \text{if } C(d_i, d_j) = \text{`Fast'}, &\allowbreak \\
   %  \quad \exists (m_i, m_j) \in L' : C(m_i, m_j) = \text{`Fast'}; &\allowbreak \\
   %  \quad \exists (m_i, m_j), (m_k, m_l) \in L' :  \\
   %  \quad \quad c(m_i, m_j) = C(m_k, m_l) = \text{`Slow'}, \\
   %  \text{if } C(d_i, d_j) = \text{`Slow'}, &\allowbreak \\
   %  \quad \exists (m_i, m_j) \in L' : C(m_i, m_j) = \text{`Slow'}. & \\
   %  \end{cases}
   %  \end{split}
   %  \normalsize  % Reset font size after equation
   %  \end{equation*}
   where \(L'_{(d_i, d_j)}\) represents the subset of SFT links \((m_i, m_j)\) that are mapped to the physical link \((d_i, d_j)\), \(R(m_i, m_j)\) represents resource requirements of the microservice link \((m_i, m_j)\), and \(C(d_i, d_j)\) denotes the maximum transmission capacity of the physical link \((d_i, d_j)\).
   
   \item \textbf{Latency Constraint.} We use the hop count to limit the data transmission latency between any pair of microservices placed in the fog devices. This constraint ensures that the number of hops does not exceed a predefined maximum value. This limitation helps to achieve deadline-oriented placement by keeping communication delays within acceptable bounds, as each additional hop can introduce a delay. We represent this constraint as:
   \begin{equation*}
    \forall (m_i, m_j) \in L': dist(d_i, d_j) \leq H_{max}
    \end{equation*}
    where $L'$ is the set of all connectivity links in the SFT, $dist(d_i, d_j)$ is a function that retrieves the number of hops between fog devices $d_i$ and $d_j$, which host microservices $m_i$ and $m_j$, respectively, and $H_{max}$ is the predefined maximum allowable number of hops for any communication path between any pair of microservices.
\end{itemize} 

% These constraints ensure a logical and efficient mapping from the SFT (microservices) to the physical network (fog devices), maintaining the integrity of data processing and communication.

In the context of this problem formulation and constraints, we explore examples of valid and invalid mappings of SFTs to physical networks. For simplicity, we assume that a device $d_i$ can have a `Big' or `Small' resource capacity. A device $d_i$ with a `Big' resource capacity can host either two microservices that require `Small' resources or one microservice that requires a `Big' resource. On the other hand, $d_i$ with a `Small' resource capacity can accommodate only one microservice that requires a `Small' resource. Furthermore, a `Fast' link in a physical network can accommodate two `Slow' links from an SFT. However, a `Slow' link in a physical network can only accommodate one `Slow' link from an SFT.

\begin{figure} [t]
    \centering
    \includegraphics[width=0.8\linewidth, trim = 4.2cm 6.2cm 8.2cm 2.2cm, clip]{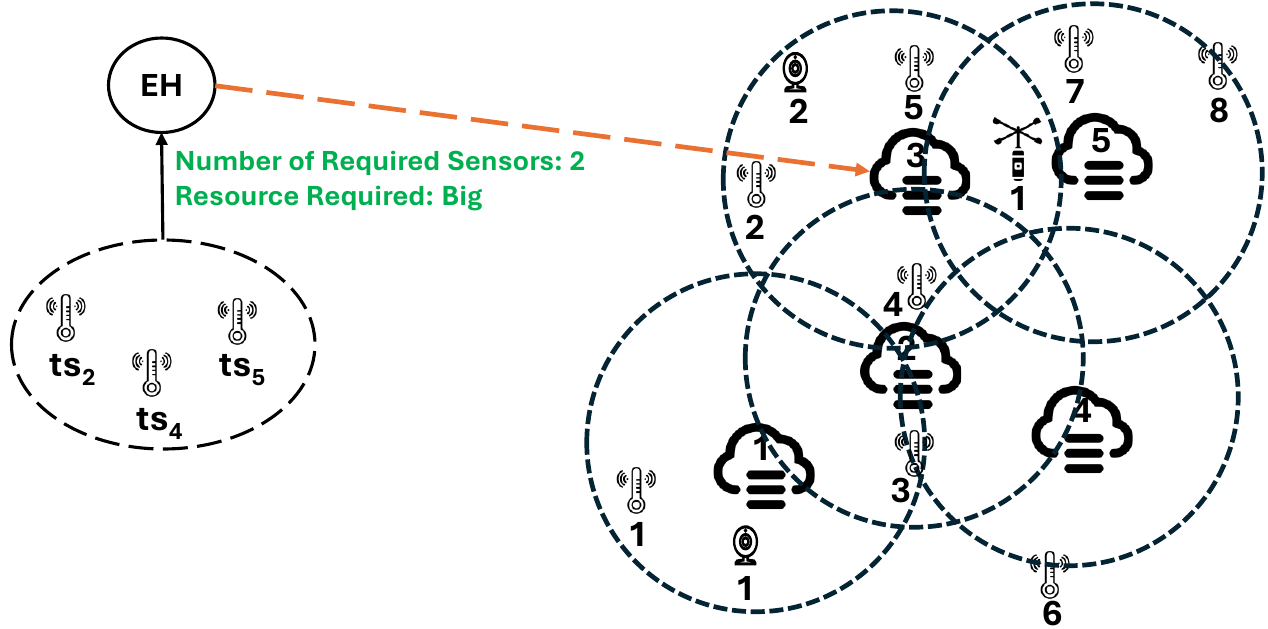}
    \caption{Valid Mapping of Simple SFT to a Physical Network}
    \label{simpleValidMapping}
\end{figure}

We first consider a simple SFT with one microservice (i.e., service function) \(M = \{m_1\}\), where $m_1$ denotes an \(EH\) requiring resource capacity of \(R(m_1) = \text{Big}\), and access to any two temperature sensors from set $\{t_{s_2}, t_{s_4}, t_{s_5}\}$. We consider a physical network with five fog devices \(D = \{d_1, d_2, d_3, d_4, d_5\}\) having capacities \( C = \text{\{Big, Small, Big, Small, Big\}}\), respectively, and with various sensors in their communication range. One possible valid mapping is the allocation of microservice $m_1$ to fog device $d_3$, which respects the sensor selection and microservice mapping and allocation constraints. Fig. \ref{simpleValidMapping} illustrates this valid mapping of the SFT onto the physical network.

\begin{figure} [t]
% trim=left bottom right top
    \centering
    \includegraphics[width=\linewidth, trim = 0cm 0.5cm 0cm 1.5cm, clip]{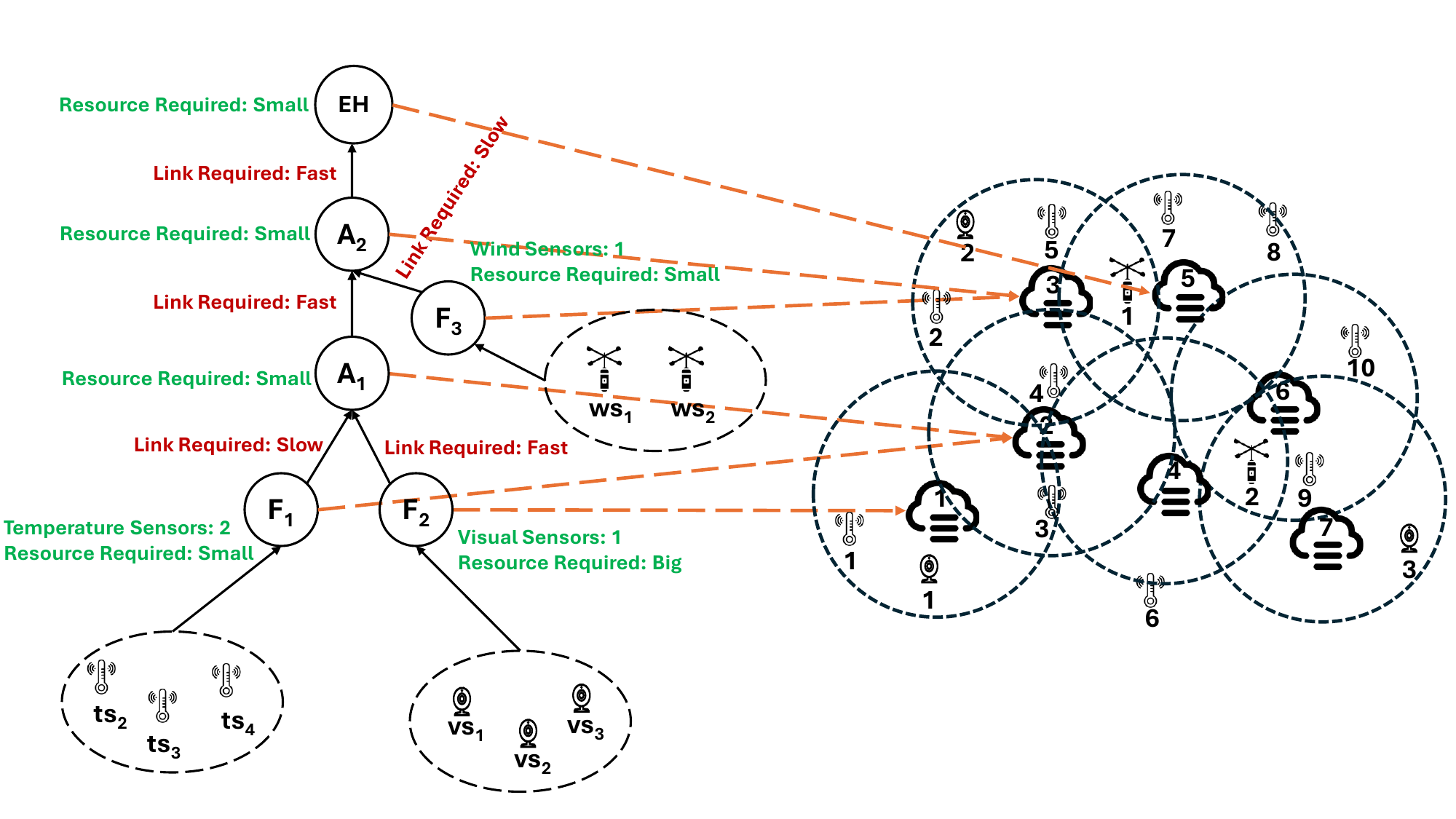}
    \caption{Valid Mapping of Complex SFT to a Physical Network}
    \label{complexValidMapping}
\end{figure}

Now consider a more complex SFT with six microservices \(M = \{m_1, m_2,\ldots, m_6\}\), where $m_1$, $m_2$, and $m_4$ denote filters \(F_1\), \(F_2\), and \(F_3\), $m_3$ and $m_5$ denote aggregators \(A_1\) and \(A_2\), and $m_6$ denotes an event handler \(EH\), respectively. The resource requirements of each microservice are \(R = \text{\{Small, Big, Small, Small, Small, Big\}}\), respectively. The connectivity links in the SFT require link speeds of \(L' = \{\text{Slow, Fast, Fast, Slow, Fast}\}\), respectively. Filter~$F_1$ requires access to two temperature sensors and $F_2$ requires one visual sensor, chosen from the sets shown at the bottom of the figure. We consider a physical network with seven devices \(D = \{d_1, d_2, d_3, d_4, d_5, d_6, d_7\}\) having a set of capacities \( C = \text{\{Big, Big, Big, Small, Big, Big, Small\}}\), respectively. Furthermore, all communication links in the physical network are `Fast'. One possible valid mapping is illustrated in Fig. \ref{complexValidMapping}, where microservices are mapped to devices as follows: \(m_1 \text{ to } d_2\), \(m_2 \text{ to } d_1\), \(m_3 \text{ to } d_2\), \(m_4 \text{ to } d_3\), \(m_5 \text{ to } d_3\), and \(m_6 \text{ to } d_5\), respecting all the aforementioned constraints.

\begin{figure} [t]
    \centering
    \includegraphics[width=\linewidth, trim = 0.5cm 1.5cm 0.2cm 2.5cm, clip]{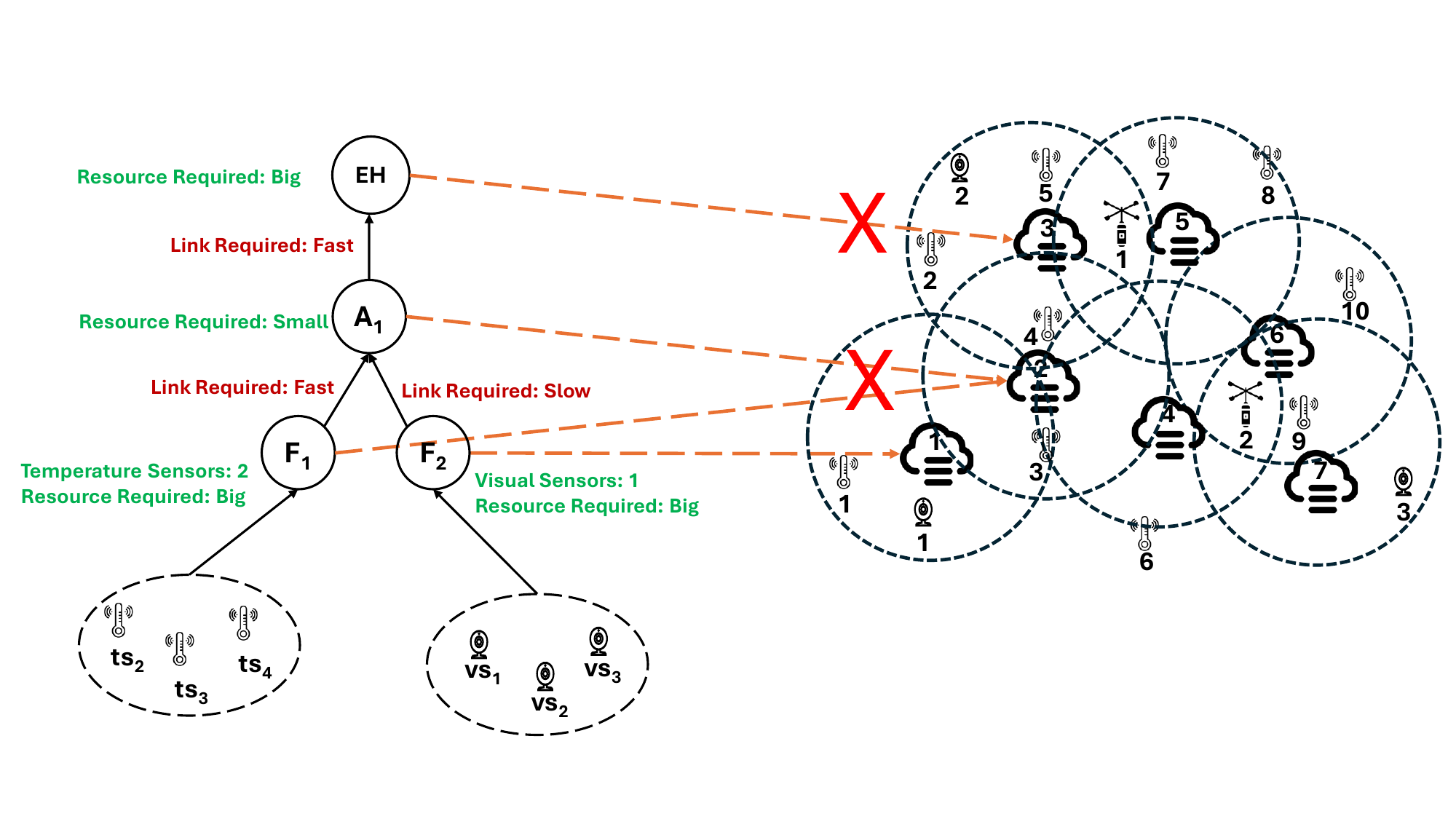}
    \caption{Invalid Mapping of an SFT to a Physical Network}
    \label{inValidMapping}
\end{figure}

Finally, consider an SFT with four microservices \(M = \{m_1, m_2, m_3, m_4\}\), where $m_1$ and $m_2$ denote \(F_1\) and \(F_2\), $m_3$ denotes an \(A_1\), and $m_4$ denotes an \(EH\), respectively. The resource requirements of each microservice are \(R = \{\text{Big, Big, Small, Big}\}\), and the filter services' sensor requirements are shown at the bottom of the tree. The link speed requirements for each connectivity link in the SFT are \(L' = \{\text{Slow, Slow, Fast}\}\). We consider a physical network with fog devices \(D = \{d_1, d_2, d_3, d_4, d_5, d_6, d_7\}\) having a set of capacities \( C = \text{\{Big, Big, Small, Small, Big, Small, Big\}}\), respectively. Furthermore, all communication links in the physical network are `Slow'. The particular mapping shown in Fig.~\ref{inValidMapping} is \emph{not} valid because it violates the resource allocation constraint by placing both a Big ($m_1$) and a Small ($m_3$) microservice on the same fog device $d_2$, which cannot host both.
%as a `Big' resource capacity fog device can only host either two `Small' resource microservices or one `Big' resource microservice, but not both.
Additionally, it violates the link capacity constraint because the link between $m_3$ and $m_4$ requires a `Fast' link, but the physical link between $d_2$ and $d_3$ is `Slow'.

\section{Service Function Tree Mapping Algorithm}

This section describes our new SFT mapping algorithm, designed to map an SFT onto a physical network while respecting all relevant constraints (i.e., sensor selection, resource allocation, and link capacity). Our algorithm adopts a bottom-up approach, prioritising the SFT's leaf microservice nodes having sensors in their ROIs. The algorithm employs backtracking and extended search mechanisms to handle situations where the initial mapping of microservice attempts to violate constraints. Our algorithm takes into account the sensor coverage of each fog device and resource limitations to find suitable placement of microservices. Additionally, it tracks link usage throughout the process to ensure the physical network's link capacity is not exceeded. Furthermore, a latency constraint is incorporated using the hop count to limit data transmission delay between microservices.
% The algorithm's output includes either a mapping of microservices to fog devices and the updated link usage information or an empty mapping and the current link usage if no valid mapping can be found.

\begin{algorithm}[t]
\caption{Map Service Function Tree to Physical Network}\label{alg1}

\begin{algorithmic}[1]
\Require
\Statex $SFT = (M, L')$, $PN = (D, L)$

\Ensure
\Statex $map: SFT \to PN$
\Function{map\_SFT\_to\_PN}{$M'$, $PN$, $SFT$, $map$, $lnkUsge$, $excDs$, $sIdx = 0$}

    \If{$excDs$ is empty}
        % \State $excDs \gets \{(m, d) \mid m \in M' \wedge d = \emptyset \}$
        \State $excDs \gets \{(m, \emptyset) \mid m \in M'\}$ %\Comment{Initialise $excDs$ with each $m$ mapped to an empty set of fog nodes}
    \EndIf
    
    \If{$sIdx \geq |M'|$}
        \State \Return $True$, $map$, $lnkUsge$
    \EndIf
    
    \State $m \gets M'[sIdx]$, $sCovDs \gets \emptyset$
    
    \For{each $d \in D$}
        \If{$d$ not in $excDs(m)$}
        
            \State  $lnkUsge_{c_1} \gets lnkUsge$
            \State $isValid, hSens, lnkUsge_{c_2} \gets$ \Call{map\_ms\_to\_fog }{$m$, $d$, $PN$, $SFT$, $map$, $lnkUsge_{c_1}$}
            \If{$hSens$}
                \State $sCovDs \gets sCovDs \cup \{hSens\}$
            \EndIf
            \If {not $isValid$ and $d$ is last in $D$ and $reqSens(m)$ and $sCovDs \neq \emptyset$}
                    \State $d_n$, $isVal$, $dSel$, $pth$, $lnkUsge_{c_2}$ $\gets$ \Call{extended\_se-arch\_to\_map}{$m$, $PN$, $SFT$, $lnkUsge_{c_1}$, $sCovDs$}
                    \If{$isVal$}
                        \State $d \gets d_n$, $isValid \gets True$
                    \EndIf
                \EndIf
            \If{not $isValid$}
                \State \textbf{Add} $d$ to $excDs(m)$ and \textbf{continue}
            \EndIf
            \State $lnkUsge_{c_1} \gets lnkUsge_{c_2}$, $map(m) \gets d$
            \State \textbf{Update} available capacity of $d$ for $m$ in $PN$
            \State $success, fiMap, lnkUsge_{c_1} \gets$ \Call{map\_SFT\_to\_PN }{$M'$, $PN$, $SFT$, $map$, $lnkUsge_{c_1}$, $excDs$, $sIdx + 1$}
                
            \If{$success$}
                \State \Return $True$, $fiMap$, $lnkUsge_{c_1}$
            \Else
                \State \textbf{Remove} $m$ from $map$ and \textbf{revert} changes to the capacity of $d$ in $PN$
                 \For{each $m' \in M'$[$sIdx + 1:$]}
                \State \textbf{Reset} $excDs(m')$
                \EndFor
            \EndIf
        \EndIf
    \EndFor
    \State \Return $False$, $\emptyset$, $lnkUsge$ %\Comment{Backtracking step if placement fails}
\EndFunction
\end{algorithmic}
\end{algorithm}

Algorithm~\ref{alg1} describes the details of our recursive backtracking algorithm to find a valid mapping of an SFT onto a physical network. The algorithm's output is a valid mapping $map$ from microservices in the SFT to fog devices in the physical network if one exists. The input is a service function tree $SFT = (M, L')$ comprising microservices $M$ and their functional dependencies $L'$ and a physical network $PN = (D, L)$ comprising fog nodes $D$ and their links $L$. We define a function $\Call{map\_SFT\_to\_PN}{}$ with parameters: $M'$ representing the list of unique identifiers derived from set $M$ of microservices, $PN$ the physical network, $SFT$ the service function tree, $map$ the current mapping of microservices, $lnkUsge$ the current usage of the links in the physical network, $excDs$ the key-value mapping of excluded fog devices for each microservice, and $sIdx = 0$ representing the default index in the list of microservice IDs (Line~1). When the algorithm is initially executed, $map$, $lnkUsge$, and $excDs$ are passed as empty mappings of key-value pairs. If $excDs$ is empty, it is initialised such that each $m \in M'$ is mapped to an empty set of fog devices, indicating that no fog devices are excluded initially (Lines~2--3). If all microservices have been successfully mapped, the algorithm returns `True', along with the current mapping $map$ and link usage information $lnkUsge$ (Lines~4--5). This is the base case of our recursive mapping algorithm. The current microservice ID $m$ is selected, and an empty set $sCovDs$ is initialised to keep track of fog devices that can cover the sensors required by the microservice $m$ (Line~6). The algorithm iterates over each fog device $d$ in $D$. If $d$ is not in the excluded fog devices $excDs$ for $m$, the algorithm attempts to map $m$ to $d$ (Lines~7--8). In Line~9, the algorithm copies the link usage to $lnkUsge_{c_1}$ to avoid modifying the original link usage information in case of backtracking. It attempts to map $m$ to $d$ by calling the function $\Call{map\_ms\_to\_fog}{}$ (Line~10). This function returns $isValid$ to indicate whether the mapping is valid, $hSens$ to indicate whether the device covers required sensors, and $lnkUsge_{c_2}$ for the updated link usage if the mapping goes through.

Algorithm~\ref{alg2} describes the details of the function $\Call{map\_ms\_to\_fog}{}$. If $hSens$ is true, then $d$ is added to the set of sensor-covering fog devices $sCovDs$ (Lines~11--12). If the mapping is invalid, the algorithm checks if $d$ is the last fog device, if $m$ requires some sensors, and if there are devices that can cover the required sensors. An extended search is then performed by calling the function $\Call{extended\_search\_to\_map}{}$ to find a suitable mapping (Lines~13--14). This function returns $d_n$ indicating a suitable fog device to host $m$, $isVal$ indicating the placement result, $dSel$ indicating the sensor-connected fog device, $pth$ representing the path between $d$ and $d_n$, and $lnkUsge_{c_2}$ indicating the updated link usage information. If the extended search is successful, $d$ is updated with the new suitable device $d_n$ and $isValid$ is set to `True' (Lines~15--16). If the mapping is still invalid, the device $d$ is added to the exclusion set in $excDs$ for $m$, and the algorithm continues to the next device for $m$ (Lines~17--18). If a valid mapping is found, the link usage is updated, $d$ is added to the map for $m$, and the available capacity of $d$ in $PN$ is updated to reflect the resource usage of $m$ (Lines~19--20). The function $\Call{map\_SFT\_to\_PN}{}$ is recursively called to place the next microservice in the $SFT$ (with index $sIdx + 1$) (Line~21). If all microservices are placed successfully in the recursive call, the function returns `True' along with the final mapping and updated link usage (Lines~22--23). If no valid mapping is found for the current microservice $m$ after iterating over all devices in $D$, the function returns `False', along with an empty map and link usage (Line~28). This triggers a backtracking process that involves removing the last successfully placed \(m\) from \(map\), reverting changes to the capacity of \(d\) in \(PN\), and resetting \(excDs\) for each subsequent microservice \(m\) (Lines~24--27). After backtracking, the algorithm searches for a valid mapping by considering alternative devices not previously explored. This continues until a valid mapping is found or all devices are fully explored for SFT mapping.

\begin{algorithm}[t]
\caption{Map Microservice to the Fog Device}\label{alg2}
\begin{algorithmic}[1]
\Function{map\_ms\_to\_fog}{$m$, $d$, $PN$, $SFT$, $map$, $lnkUsge_{c_1}$}
    \State $hSens \gets False$
    \If{\Call{has\_required\_sensors}{$d$, $reqSens(m)$, $reqSens\newline Cnt(m)$}}
        \If{$reqSensCnt(m) \neq 0$}
            \State $hSens \gets True$
        \EndIf
        \If{\Call{is\_resource\_compatible}{$d$, $reqCap(m)$, $PN$, $SFT$}}
            \State $preds \gets$ \Call{get\_predecessor\_microservices}{$m$}
            \If{$preds = \emptyset$}
                \State \Return $True$, $hSens$, $lnkUsge_{c_1}$
            \EndIf
            \If{$map$ is not empty}
                \State $sameDPreds \gets \{p \in preds \,|\, map(p) = d\}$
                \State $otherDPreds \gets preds \setminus sameDPreds$
                
                \If{$sameDPreds \neq \emptyset$}
                    \For{each $p \in otherDPreds$}
                        \State $d_p \gets map(p)$
                        \State $isVal$, $lnkUsge_{c_2} \gets$ \Call{validate\_path}{$PN$, $SFT$, $p$, $m$, $d_p$, $d$, $lnkUsge_{c_1}$}
                        \If{not $isVal$}
                            \State \Return $False$, $hSens$, $lnkUsge_{c_1}$
                        \Else
                            \State $lnkUsge_{c_1} \gets lnkUsge_{c_2}$
                        \EndIf
                    \EndFor
                    \State \Return $True$, $hSens$, $lnkUsge_{c_1}$
                \EndIf
                \State $isVal$, $lnkUsge_{c_2} \gets$ \Call{validate\_connectivity\_ and\_link\_speed}{$PN$, $SFT$, $m$, $d$, $map$, $lnkUsge_{c_1}$}
                \If{$isVal$}
                    \State $lnkUsge_{c_1} \gets lnkUsge_{c_2}$
                    \State \Return $True$, $hSens$, $lnkUsge_{c_1}$
                \EndIf
            \EndIf
        \EndIf
    \EndIf
    \State \Return $False$, $hSens$, $lnkUsge_{c_1}$
\EndFunction
\end{algorithmic}
\end{algorithm}

\subsection{Single Microservice Mapping}

Algorithm~\ref{alg2} determines whether a given microservice $m$ can be mapped to a specific fog device $d$ considering sensor requirements, resource compatibility, and connectivity and latency constraints. The function \Call{map\_ms\_to\_fog}{} takes six parameters: the microservice $m$, the fog device $d$, the physical network $PN$, the service function tree $SFT$, the current mapping of microservices to fog devices $map$, and the current link usage $lnkUsge$ (Line~1). A boolean variable $hSens$ (has required sensors) is initialised as `False' (Line~2). The algorithm checks if the fog device $d$ possesses the required sensors for the microservice $m$ using \Call{has\_required\_sensors}{} (Line~3). If $d$ has required sensors and the count of required sensors is not zero, $hSens$ is set to `True' (Lines 4--5). This condition ensures that only those fog devices with required sensors in their range are added to the $sCovDs$ (described in Algorithm~\ref{alg1}). In Line~6, a function \Call{is\_resource\_compatible}{} is used to find if the fog device $d$ has the resources required to host the microservice $m$. If the fog device is resource-compatible, the function \Call{get\_predecessor\_microservices}{} is called to get the predecessor microservices of $m$ (Line~7). If there are no predecessor microservices, the function returns `True' to indicate successful mapping, $hSens$ for the sensor availability, and $lnkUsge_{c_1}$ for the current link usage (Lines~8--9). However, suppose this microservice has predecessor microservices, and a mapping exists (i.e., $map$ is not empty). In that case, the algorithm separates predecessors of `$m$` into two sets: (`$sameDPreds$`) which are mapped to the same device and (`$otherDPreds$`) which are mapped to other devices (Lines~10--12). If there are predecessors mapped to the same device, the algorithm iterates through each predecessor mapped to other devices. It validates the path between the predecessor's device and `$d$` for the microservice $m$ using \Call{validate\_path}{}, which validates connectivity and latency constraints. If the path is invalid, the function returns `False', $hSens$ status, and $lnkUsge_{c_1}$. If the path is valid, $lnkUsge_{c_1}$ is updated with $lnkUsge_{c_2}$. After iterating over all predecessor's devices, the function returns `True', $hSens$, and $lnkUsge_{c_1}$ (Lines~13--21). The path connectivity, latency, and link requirements are not verified when directly connected microservices are mapped to the same fog device. If none of the predecessors for $m$ exists in the same fog device, then the algorithm validates the overall connectivity and link speed requirements for mapping $m$ to $d$ using the function \Call{validate\_connectivity\_and\_link\_speed}{} (Line~22). The algorithm returns `True', $hSens$ status, and updated link usage $lnkUsge_{c_1}$ if the validation is successful (Lines~23--25). Otherwise, the algorithm returns `False', $hSens$, and the original link usage $lnkUsge_{c_1}$ indicating that microservice $m$ cannot be mapped to device $d$ (Line~26).

\begin{algorithm}[t]

\caption{Extended Search for Mapping}\label{alg3}
\begin{algorithmic}[1]
\Function{extended\_search\_to\_map}{$m$, $PN$, $SFT$, $lnkUsge_{c_1}$, $sCovDs$}
    \State $d_n \gets$ Null, $dSel \gets$ Null
    \State $pth \gets \emptyset$, $minHops \gets +\infty$
    \State Initialise $lnkUsge_{c_2}$ as an empty mapping
    \State $linkReq \gets$ \Call{det\_link\_requirement}{$m$}
    \For{each $sd \in sCovDs$}
        \State $visited \gets \emptyset$, $queue \gets \{(sd, 0, \emptyset)\}$
        \While{$queue$ is not empty}
            \State $cd$, $distance$, $curPth$ $\gets$ \Call{Dequeue}{$queue$}
            \If{$cd \notin visited$}
                \State $visited \gets visited \cup \{cd\}$
                \If{\Call{is\_resource\_compatible}{$cd$, $reqCap(m)$, $PN$, $SFT$}}
                    \State $val\_link$, $lnkUsge_{c_3}$ $\gets$ \Call{validate\_extended\_\newline path}{$PN$, $\{sd\} \cup curPth$, $linkReq$, $lnkUsge_{c_1}$}
                    \If{$val\_link$ and $distance$ $<$ $minHops$}
                        \If{$cd \notin sCovDs$}
                            \State $d_n \gets cd$, $dSel \gets sd$
                            \State $pth \gets curPth$, $minHops \gets distance$
                            \State $lnkUsge_{c_2} \gets lnkUsge_{c_3}$
                            \State \textbf{break}
                        \EndIf
                    \EndIf
                \EndIf
                \For{each $neighbor \in$ \Call{neighbors}{$cd$, $PN$}}
                    \If{$neighbor \notin visited$}
                        \State $queue \gets queue \cup \{(neighbor$, $distance$ + 1, $curPth \cup \{neighbor\})\}$
                    \EndIf
                \EndFor
            \EndIf
        \EndWhile
    \EndFor
    \If{$d_n \neq$ Null}
        \State $lnkUsge_{c_1} \gets lnkUsge_{c_2}$
        \State \Return $d_n$, $True$, $dSel$, $pth$, $lnkUsge_{c_1}$
    \Else
        \State \Return Null, $False$, Null, $\emptyset$, $lnkUsge_{c_1}$
    \EndIf
\EndFunction
\end{algorithmic}
\end{algorithm}

\subsection{Extended Search for Mapping}

Algorithm~\ref{alg3} employs a breadth-first search (BFS) strategy to perform an extended search to map a given microservice $m$ to a fog device with the minimum number of hops from the sensor-connected fog device, satisfying the resource and link requirements. The function \Call{extended\_search\_to\_map}{} takes five parameters: the microservice $m$, the physical network $PN$, the service function tree $SFT$, the current link usage $lnkUsge_{c_1}$, and the set of sensor-connected fog devices $sCovDs$ (Line~1). Several variables are initialised, and the link requirements are determined based on the type of sensors in the ROI of the microservice $m$ using the function \Call{det\_link\_requirement}{} (Lines~2--5). The algorithm then iterates through each sensor-connected fog device $sd$ in $sCovDs$ and initialises a set $visited$ and a queue $queue$ with the current $sd$, its distance $(0)$, and an empty path for each $sd$ (Lines~6--7). A BFS strategy is then employed to explore all possible paths starting from each $sd$ and iterating through all its neighbours (Lines~8--22). This strategy involves validating the resource, link, and latency requirement constraints and then finding a fog device $d_n$ with the minimum number of hops to $sd$. Upon finding a suitable new fog device, the algorithm proceeds to update the current link usage with the new link usage $lnkUsge_{c_2}$. It then returns the following values: $d_n$ representing the closest fog device, $True$ representing the successful mapping of the microservice $m$, $dSel$ representing the sensor-connected fog device, $pth$ representing the path between $sd$ and $d_n$, and $lnkUsge$ reflecting the updated link usage after microservice placement (Lines~23--25). If the extended search cannot find a suitable fog device, the algorithm returns a failure flag along with Null or default values for other return variables (Lines~26--27). The remaining functions and their implementation details are available \href{https://drive.google.com/drive/folders/1VvrxfucTL27ac6IgXcMiJwC_XqPC_ord?usp=drive_link}{\textcolor{blue}{elsewhere}}.

Here we analyse the time and space complexity of our algorithms. Algorithm~\ref{alg1} has a time complexity of $O(m^n)$ due to the exhaustive search for $n$ microservices across $m$ devices. Algorithm~\ref{alg2} has a $O(p \times l)$ complexity due to evaluating $p$ predecessors with average $l$ links per predecessor for mapping. In Algorithm~\ref{alg3}, a BFS search is employed to $s$ devices, which results in a time complexity of $O(s \times (V + E))$, where $V$ and $E$ represent all devices and their physical links. The space complexity of Algorithm~\ref{alg1} is $O(M \times D + M + N)$, considering potential mappings and network representation. Algorithm~\ref{alg2} has a space complexity of $O(P + L)$, considering predecessors and link usage. Finally, Algorithm~\ref{alg3} has a space complexity of $O(V + E)$, considering network exploration. 

\section{Practical Evaluation}

We evaluated the effectiveness of our new algorithm by deploying it on various scenarios ranging from simple to complex. Digital construction and IIoT research validation are largely simulation-based due to the challenges of deploying a large-scale network of sensors and fog devices in a real-world IIoT site~\cite{svorobej2019simulating,yeung2022role}. Therefore, we adopted a simulation approach for our experiments that leverages existing case studies on sensor types and coverage in concrete pouring~\cite{arabshahi2021review} to create a realistic and controlled environment.

\subsection{Experiment Settings with Augmented Datasets}

We use the NetworkX Python library to build a physical network of fog devices, simulating an industrial building's floor slab ranging from 100 to 400 sqm. This space is segmented into evenly divided sections (e.g., 10 sqm) for concrete pouring. A specific area, designated as the ROI, focuses on sensor coverage for effective real-time data processing. Event processing aims to detect drying and structural defects, each requiring different sensors. Drying involves temperature, wind, humidity, and ROI-attached moisture sensors, while structural involves formwork visual, strain gauge, and movement sensors. However, for these experiments, we use temperature and wind sensors to monitor the drying process and visual sensors to identify structural defects. Datasets are generated based on realistic assumptions about sensor types and their coverage in IIoT sites. The simulation considers varying data volumes from different sensors to reflect actual data processing demands. For instance, temperature sensors generate data more frequently than strain gauges, which monitor rebar stress and may not require continuous monitoring.

In our setup, fog devices with fixed resource capacities (`Big' or `Small') are interconnected through various communication links, each with a defined speed (`Fast' or `Slow'). Datasets for physical networks with up to 10 devices are generated based on realistic assumptions about sensor and fog device coverage in IIoT sites. For larger networks, datasets are created following specific rules to ensure connectivity and sensor coverage across devices. Some of the rules are as follows: (1) each device communicates with at least one neighbour device, (2) each device has at least one sensor in its range, (3) devices can have multiple sensor types, and (4) some sensors require coverage by multiple devices. We define the SFT where each node represents a microservice with specific resource requirements (`Big' or `Small') and is connected to other microservices requiring specific link speeds (`Fast' or `Slow'). Leaf nodes represent microservices requiring sensor data within their designated ROIs. The structure of each SFT is predefined, ranging from the simple with a single SFC to more complex structures with multiple interconnected SFCs. All algorithms are written in Python with experiments conducted on an Intel Core i7-1255U processor (1.70 GHz and 16.0 GB RAM). Table \ref{table:var} provides the details of experimental variables.

% Table 3.1: Event Processing Sensors
% \begin{center} \caption{Sensor Types for Drying and Structural Defect Detection} \label{table:event_processing_sensors} \begin{tabular}{|l|c|}
% \hline \textbf{Event Processing Goal} & \textbf{Required Sensor Types} \\
% \hline Drying Detection & Temperature, Wind, Unattached Moisture, RIO Attached Moisture \\
% \hline Structural Defect Detection & Various RIO and Formwork Visual, Structural, Movement Sensors \\
% \hline
% \end{tabular}
% \end{center}

\begin{table}[t]
\centering
\caption{Experimental Variables}
\begin{tabular}{|l|c|}
\hline
\textbf{Variable Name}&{\textbf{Value}} \\
\hline
Number of Fog Devices in Physical Network &  [5, 25]\\
Number of Temperature Sensors & [8, 35]\\
Number of Visual Sensors & [2, 11]\\
Number of Wind Sensors & [1, 5]\\
Number of Microservices in SFT & [3, 11]\\
\hline
\end{tabular}
\label{table:var}
\end{table}

\begin{figure*}[ht]
    \centering
    \begin{subfigure}[b]{0.3\linewidth}
        \includegraphics[width=\linewidth]{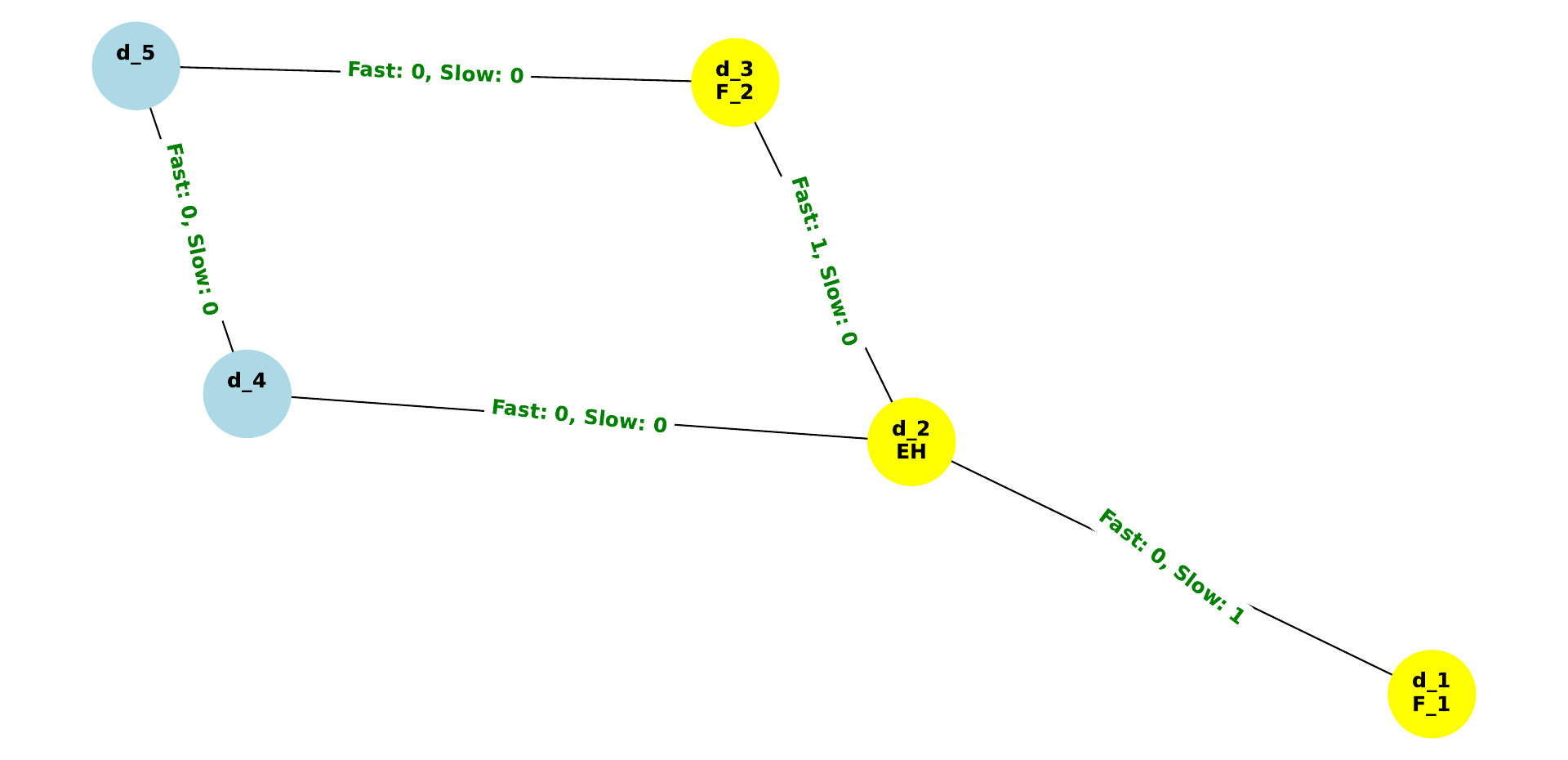}
        \caption{PN: 5 Fog Devices, SFT: 3 Microservices}
        \label{fig:result1}
    \end{subfigure}
    \hfill % This inserts a horizontal space between subfigures
    \begin{subfigure}[b]{0.30\linewidth}
        \includegraphics[width=\linewidth]{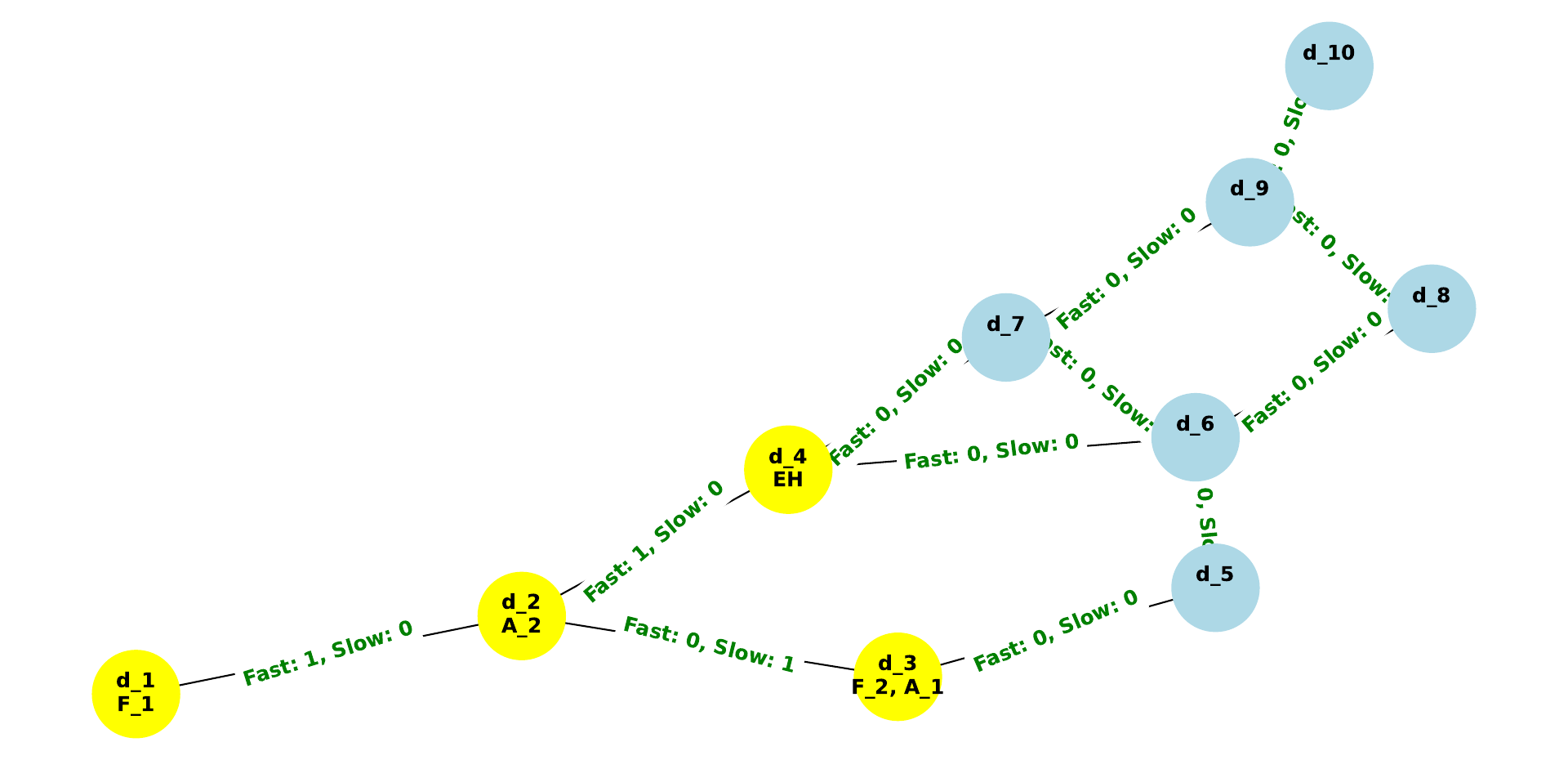}
        \caption{PN: 10 Fog Devices, SFT: 5 Microservices}
        \label{fig:result2}
    \end{subfigure}
    \hfill
    \begin{subfigure}[b]{0.30\linewidth}
        \includegraphics[width=\linewidth]{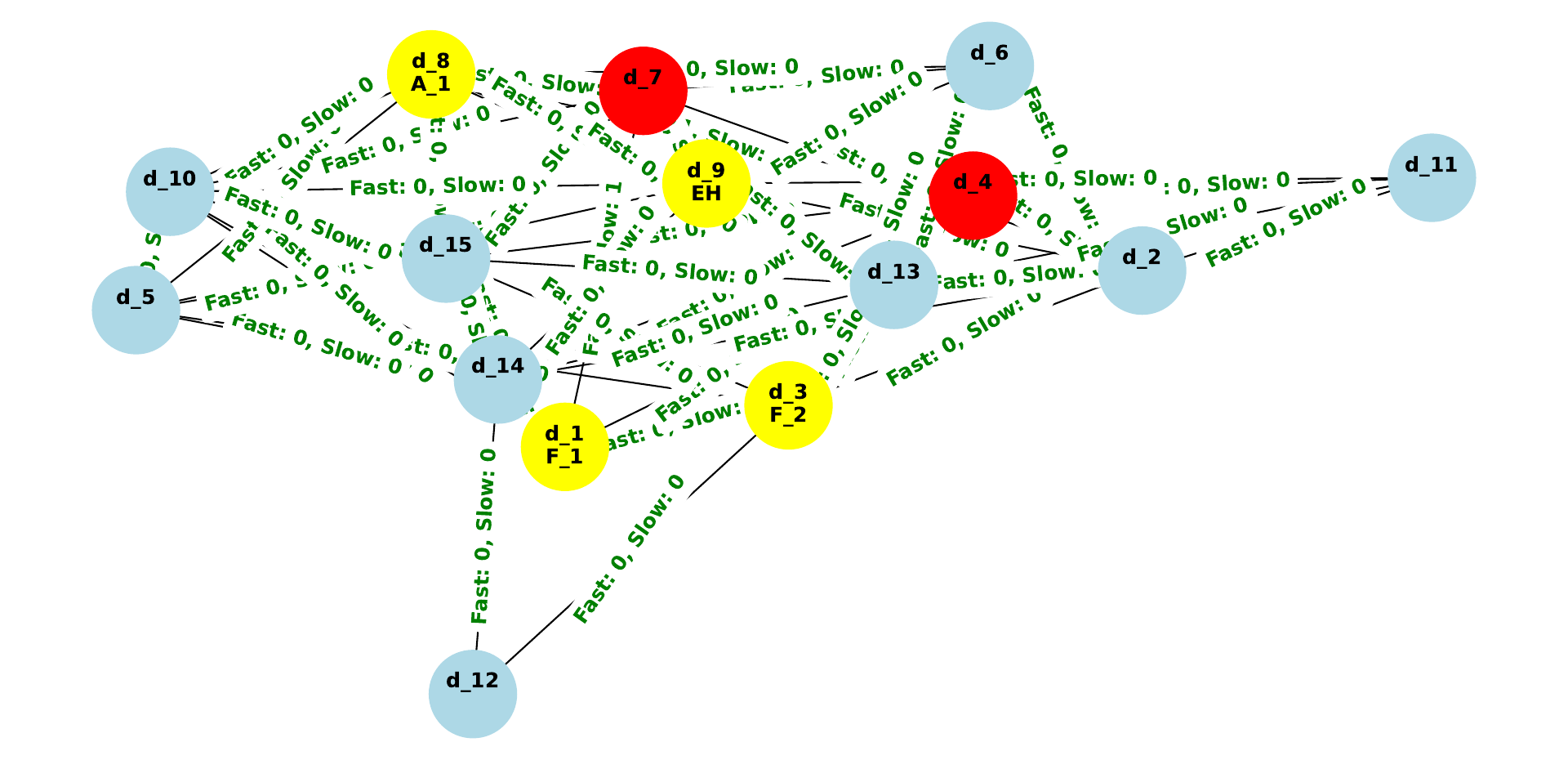}
        \caption{PN: 15 Fog Devices, SFT: 4 Microservices}
        \label{fig:result3}
    \end{subfigure}
    \par\bigskip % This adds some vertical space before the next row of subfigures
    \begin{subfigure}[b]{0.32\linewidth}
        \includegraphics[width=\linewidth]{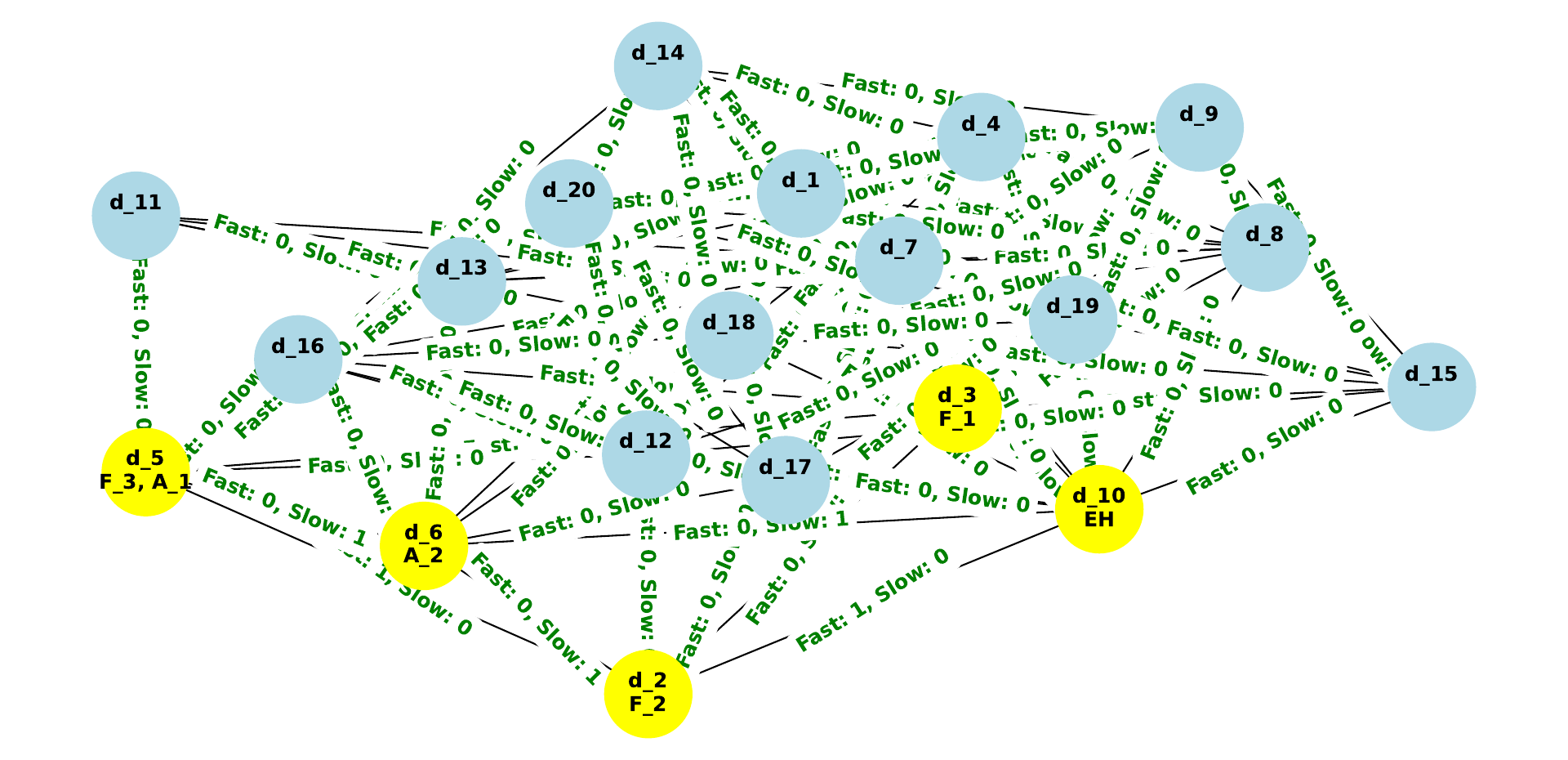}
        \caption{PN: 20 Fog Devices, SFT: 6 Microservices}
        \label{fig:result4}
    \end{subfigure}
    \hfill
    \begin{subfigure}[b]{0.33\linewidth}
        \includegraphics[width=\linewidth]{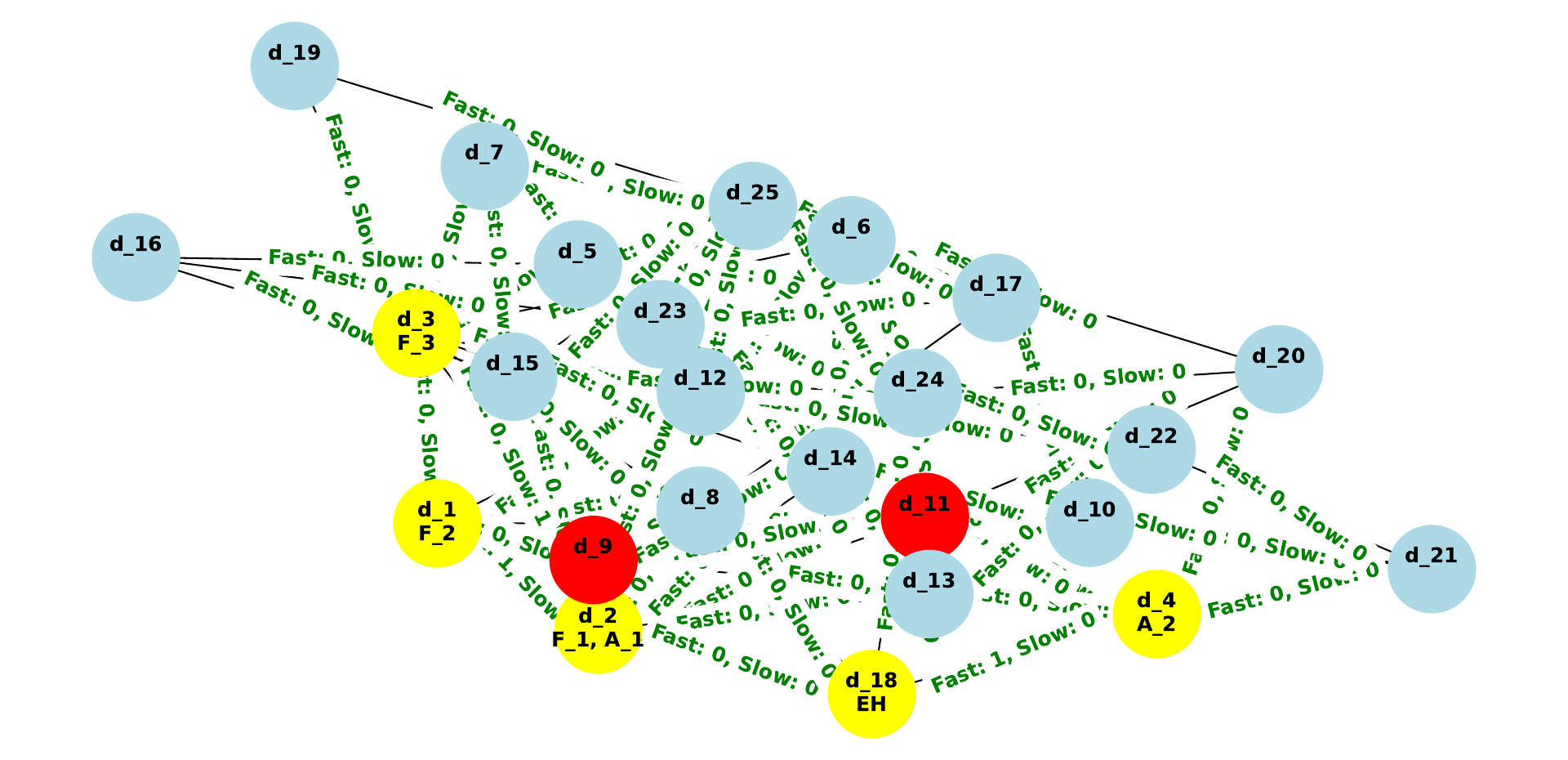}
        \caption{PN: 25 Fog Devices, SFT: 6 Microservices}
        \label{fig:result5}
    \end{subfigure}
    \hfill
    \begin{subfigure}[b]{0.33\linewidth}
        \includegraphics[width=\linewidth]{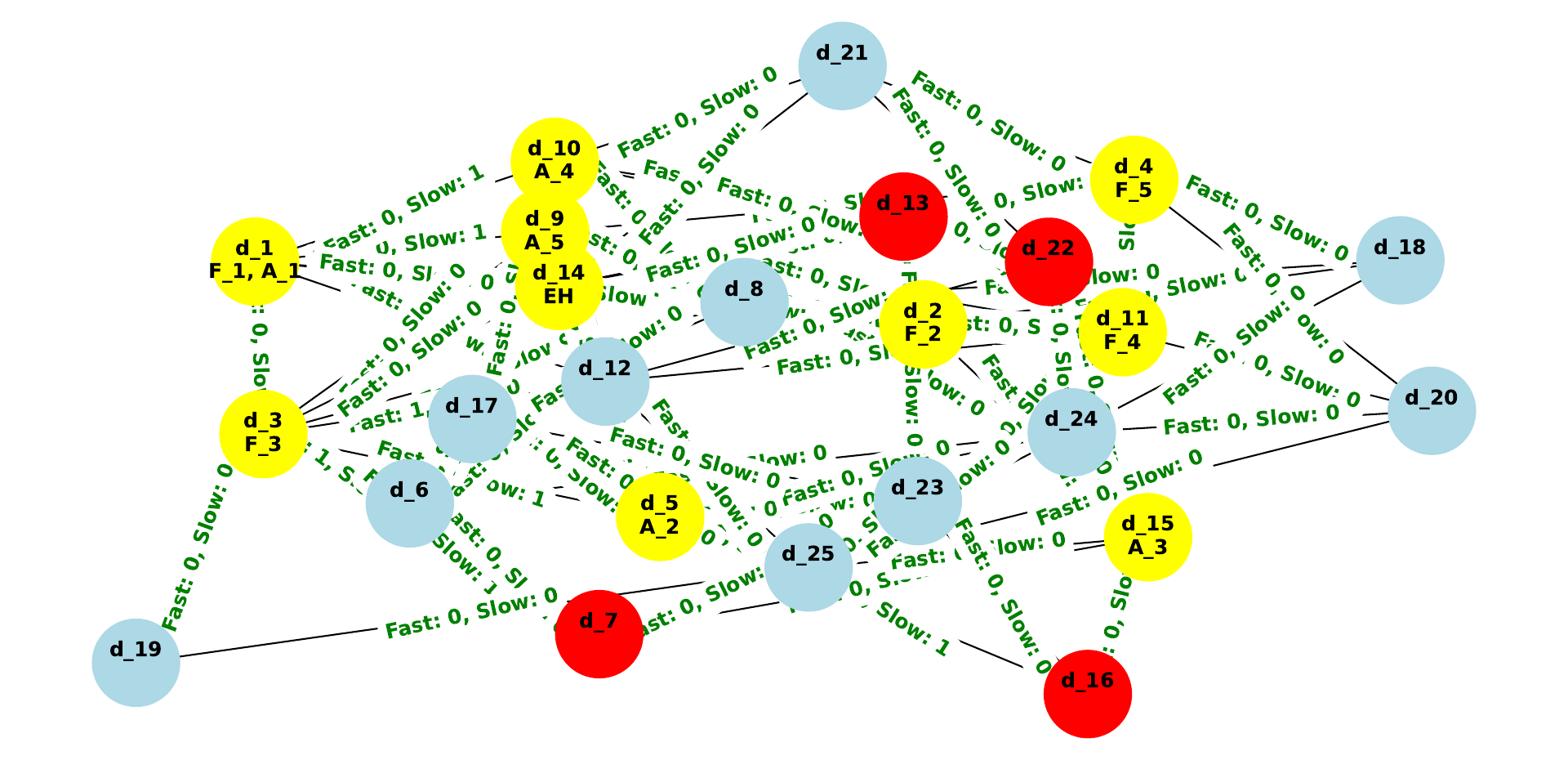}
        \caption{PN: 25 Fog Devices, SFT: 11 Microservices}
        \label{fig:result6}
    \end{subfigure}
    
    \caption{SFT mapping onto Physical Networks}
    \label{fig:networks}
    \vspace{-0.4cm}
\end{figure*}

\subsection{Discussion}

Our SFT mapping algorithm facilitates real-time event processing within an IIoT site focused on concrete pouring by mapping microservices to suitable devices. This mapping aims to support detecting two critical events: drying and structural defects. Each event requires different sensor types to ensure timely and accurate detection. Data coverage is confined to a specific area within the IIoT site, not the entire site.

The effectiveness of our proposed SFT mapping algorithm is demonstrated through various scenarios of microservice mapping onto physical networks. Figure~\ref{fig:networks} shows varying complexities of physical network topologies and microservice mappings. For simpler structures, SFT's mappings onto physical networks, as shown in Figs.~\ref{fig:result1} and~\ref{fig:result2}, can be manually validated. However, the complexity of real-world scenarios necessitates an automated algorithmic approach. Such complexity is depicted in Figs.~\ref{fig:result3} through~\ref{fig:result6}. Our SFT mapping algorithm proves to be particularly effective in complex mappings, where manual strategies are time-consuming and prone to error. By leveraging backtracking and extended search strategies, our algorithm explores the search space and identifies valid placements for complex SFTs while respecting the constraints. It finds mappings within the specified latency based on hop count to limit data transmission delays and place microservices within the desired ROI. This highlights the practicality of our approach to finding a valid mapping.

In the figures, the light blue nodes represent the default state of each physical node. Physical nodes hosting microservices are highlighted in yellow, and links are labelled with the number of Fast and Slow connections they support. The physical nodes highlighted in red represent the data forwarding nodes, which result in the case of extended search performed during the mapping process. Our algorithm exhibits several characteristics of a valid mapping. In line with our original motivation, the mapping is performed only in specific ROIs where required sensors are located. This clustering of devices around the ROI minimises latency by enabling direct communication and avoids placing microservices in distant fog devices. In some instances, it may not be possible to map directly connected microservices in SFT onto the same or directly connected fog devices in the physical network. Our algorithm finds the intermediate nodes and labels them in red. In most cases, our algorithm finds a mapping that reduces the total number of hops. Depending on the resource and link capacities, fog devices host the maximum number of microservices possible. However, when there are too many microservices, these microservices may be scattered across the network due to resource and link limitations.

\section{Conclusion}

We have presented an SFT mapping algorithm to address the challenge of microservice placement in IIoT networks. Traditional single-task placement strategies often struggle with complex event-processing pipelines involving interdependent microservices. Our algorithm addresses these limitations by leveraging fog computing infrastructure and accounting for various constraints, such as function chain interdependencies, required microservice execution orders, and sensor availability in their regions of interest. The algorithm employs backtracking and extended search strategies to ensure valid SFT mappings under resource limitations and sensor constraints. The performance of our new algorithm was evaluated through simulations involving both simple and complex scenarios. Results demonstrate its effectiveness in facilitating seamless deployment of microservices in fog networks.
% This highlights its advantage over manual placement strategies, which can be time-consuming and error-prone for complex scenarios.

As the complexity and size of SFTs increase, the computational time of our algorithm may also increase. Future work will explore machine learning and heuristic-based optimisation techniques to reduce these computational times. Currently, leaf microservice nodes process data streams from sensors of the same modality within a single fog device's communication range. In more complex scenarios, however, leaf nodes may need to integrate data from sensors of varied modalities across multiple devices. Future work will focus on devising strategies to handle these complex situations during the mapping process.

\section*{Acknowledgment}
This research was supported in part by Australian Research Council Discovery Project: DP220101516, ‘Embedding Enterprise Systems in IoT Fog Networks Through Microservices’.

\bibliographystyle{IEEEtran}
\bibliography{references}

\end{document}